\documentclass[twocolumn,showpacs,preprintnumbers,amsmath,amssymb]{revtex4}
\usepackage{amsmath,amsfonts,diagbox,latexsym,amssymb,graphics,epsfig,subfigure,color,makeidx}
\usepackage{xcolor}
\usepackage[utf8]{inputenc}
\usepackage{graphicx,hyperref}
\usepackage{bm}
\usepackage{color}
\usepackage{multirow}

\begin{document}
	
	\title{Effects of nonlinear interactions on the superradiant instability of charged black holes}
	\author{Bo-Wen Qin,
    Yu-Peng Zhang\footnote{zhangyupeng@lzu.edu.cn, corresponding author},
	}
	\affiliation{
        Key Laboratory of Quantum Theory and Applications of MoE, Lanzhou Center for Theoretical Physics, Key Laboratory of Theoretical Physics of Gansu Province, Lanzhou University, Lanzhou 730000, China\\
        Institute of Theoretical Physics \& Research Center of Gravitation, Lanzhou University, Lanzhou 730000, China
	}
	
	\begin{abstract}

	A Reissner-Nordstr\"{o}m black hole (RNBH) enclosed in a cavity is known to be superradiantly unstable to charged scalar perturbations below a critical frequency. Inspired by the emergence of the QCD axion as a prominent dark matter candidate, we construct a model featuring an axion field coupled to an electromagnetic field that undergoes superradiant growth around an RNBH. Utilizing numerical relativity, we achieve stable, long-term evolution of this system and perform a comparative analysis across various parameter spaces. Our comprehensive investigation reveals the formation of a ``hairy" black hole, whose final state is governed by a diverse set of physical parameters. Notably, the decay constant in the axion potential—representing nonlinear interactions—bifurcates the superradiant instability into two distinct behavioral regimes, leading to more significant dynamical shifts than previously reported. Furthermore, we examine the influence of the scalar field’s charge and mass, as well as the mirror’s position. We investigate the axionic bosenova process and observe a long-term beating pattern of the axion field induced by nonlinear interactions. By fine-tuning these parameter combinations, we demonstrate that the system can evolve toward a variety of distinct physical endpoints.

	\end{abstract}
	\maketitle
\section{Introduction} \label{sec:intro}

Black holes, as extremely compact objects in the universe, have been enthusiastically studied since their existence were predicted by general relativity last century. Characterized by singularities and event horizons, black holes serve as ideal laboratories for extreme physics, providing a robust framework for testing gravitational theories and astrophysical models \cite{Barack:2018yly}. Observational evidence from gravitational-wave detections \cite{LIGOScientific:2016aoc,LIGOScientific:2016vlm,LIGOScientific:2016lio,LIGOScientific:2016sjg,LIGOScientific:2019ysc,LIGOScientific:2020stg,LIGOScientific:2017vwq,LIGOScientific:2018dkp,LIGOScientific:2018jsj,LIGOScientific:2020kqk,KAGRA:2021duu} and black hole imaging \cite{EventHorizonTelescope:2019pgp,EventHorizonTelescope:2022wkp,Falcke:1999pj,Chen:2022scf} confirms the existence of black holes and reinforces the accuracy of General Relativity's strong-field dynamics.

Given the immense energy stored within black holes—particularly in the form of rotational energy—the study of energy extraction has become a focal point of research. These investigations are highly significant as they challenge the classical notion that nothing can be retrieved from a black hole, exploring mechanisms that bypass the constraints of the event horizon. Since Penrose uncovered the mechanism for extracting a black hole's rotational energy \cite{Penrose:1969pc,Penrose:1971uk}, studies targeting the extraction of its energy, charge, and angular momentum have flourished, leading to numerous unique phenomena. A variety of Penrose processes—including the electric, repetitive, and collisional types—have been identified and studied \cite{Piran:1977dm,Denardo:1973pyo,Wagh:1989zqa,Ruffini:2024dwq,Ruffini:2024irc}. Beyond the Penrose process, alternative mechanisms such as the Blandford-Znajek process \cite{Blandford:1977ds} and magnetic reconnection \cite{Comisso:2020ykg,Wang:2022qmg} constitute integral components of the energy extraction landscape. Following the Penrose process, Zel'dovich identified wave amplification around rotating dissipative bodies \cite{Zeldovich:1971mw,Zeldovich:1972zqp}, while Teukolsky provided the essential linear perturbation framework \cite{Teukolsky:1972my}. These milestones led to the introduction of black hole superradiance and the ``black hole bomb" hypothesis in 1972 \cite{Press:1972zz,Brito:2015oca}.

Building upon these foundational theories, black hole superradiance has emerged as a pivotal mechanism. It describes a process whereby both massive and massless bosonic fields, when propagating around a rotating or charged black hole, can become trapped within its vicinity. Through such interaction, these fields extract energy, angular momentum, and charge from the black hole, culminating in an exponential amplification of the field amplitude \cite{Brito:2015oca}. The frequency of bosonic field must satisfy this following condition to trigger superradiant instability around rotating black hole \cite{Zeldovich:1971mw,Zeldovich:1972zqp}
\begin{equation}
    \omega < m\Omega,
\label{crifrerota}
\end{equation}
or charged black hole \cite{Bekenstein:1973mi},
\begin{equation}
    \omega < q\Phi_+,
\label{crifrechar}
\end{equation}
where $m$ is the azimuthal index which mark the rotation of the bosonic field, $\Omega$ stands for the angular velocity of the black hole,
where q is the charge of bosonic field, and $\Phi_+$ is the electric potential at the horizon of charged black hole.

Inspired by Teukolsky, at the beginning, most of the researches were focus on linear perturbation of bosonic field. Given the excellent works of Zel'dolvich and Teukolsky, linear instability of a bosonic field around rotating black holes were studied, revealing the important relationship between the maximal extracted energy, superradiant growth rate, and azimuthal index $m$ \cite{Zouros:1979iw,Detweiler:1980uk,Cardoso:2004nk,Dolan:2007mj,RosaJ}. While the superradiance of rotating black holes is a theoretically verified physical phenomenon, its exceptionally slow growth rate \cite{Dolan:2012yt,Witek:2012tr} remains a significant bottleneck, particularly for models requiring substantial dark matter accumulation. Although the linear instability of Kerr black holes against massive vector fields offers improved timescales \cite{Pani:2012vp}, RNBH in cavities have emerged as a compelling alternative \cite{Dolan:2015dha,Sanchis-Gual:2015lje,Sanchis-Gual:2016tcm, Herdeiro:2013pia, Hod:2012wmy,Bekenstein:1973mi}.

While the superradiant instabilities of free (massless or massive) scalar fields have been extensively studied within the framework of linear perturbation theory, recent focus has shifted toward the role of nonlinearities—originating from both self-interacting higher-order potentials and the gravitational backreaction on the spacetime metric. Beyond the linear regime, these nonlinear studies reveal compelling dynamics and significantly accelerated growth rates, even in rotating scenarios, often culminating in the formation of ``black hole bombs" and hairy black holes \cite{RosaJ,Sanchis-Gual:2014ewa,Herdeiro:2014goa,Zilhao:2015tya,Herdeiro:2016tmi,Ganchev:2017uuo,Herdeiro:2017phl}. Understanding these complex endpoints demands further numerical work. The requisite techniques for full nonlinear simulations, established in recent decades, have paved the way for substantive progress \cite{Witek:2010qc,Cardoso:2012qm,Okawa:2014nda}. Such nonlinear effects manifest in novel phenomena, most notably the dynamical transition to hairy black holes as equilibrium end-states, where the field's backreaction is no longer negligible \cite{Sanchis-Gual:2015lje,Sanchis-Gual:2016tcm,Okawa:2015fsa, Bosch:2016vcp,East:2017ovw,East:2018glu,Baryakhtar:2020gao}. Furthermore, the inclusion of nonlinear self-interactions can trigger diverse physical mechanisms around black holes that extend well beyond the scope of standard superradiant processes \cite{Zhang:2023qtn,Zhang:2025jlb}. The scope of superradiance extends beyond black holes, significant investigations have also been conducted on diverse compact objects, including Q-balls, boson stars, and exotic compact stars \cite{Saffin:2022tub,Zhou:2023sps,Gao:2023gof}. A variety of mechanisms have been proposed to trigger black hole superradiance. These include superradiance in modified gravity, effects from Hawking radiation, and contributions from photon or fermionic sources \cite{Luo:2024gqo, Jha:2022tdl, Piovella:2023aou, Dai:2023zcj,Dai:2023ewf}. Furthermore, superradiant instabilities have been studied in the context of electromagnetic (EM) scattering \cite{Karmakar:2023hlb}. Such diverse research highlights the broad range of physical processes that can lead to energy extraction from black holes.

A primary physical candidate for such processes is the axion, originally introduced as a pseudo-Goldstone boson of the Peccei-Quinn symmetry to resolve the strong CP problem \cite{Peccei:1977hh}. Now regarded as a leading dark matter candidate \cite{Marsh:2015xka}, ultra-light ``string axions" play a crucial role in the development of black hole instabilities \cite{Kodama:2011zc} and the emission of gravitational radiation \cite{Yoshino:2013ofa}, particularly through the nonlinear ``Bosenova" process triggered by their self-interactions \cite{Yoshino:2015nsa, Yoshino:2012kn, Blas:2020nbs}. Despite extensive investigations in Kerr spacetimes, the interplay between axionic fields and the electromagnetic structure of charged black holes presents a rich and distinct landscape. In such environments, the coupling between the axion and the electromagnetic field can trigger specialized superradiant mechanisms that differ significantly from the purely rotational case, necessitating a detailed exploration of axionic instability in RNBH backgrounds.

The growing interest in axions and axion-like particles is driven not only by their unique properties but also by their superradiant growth around rotating black holes. When these fields are coupled with black holes, they can create hairy black holes, where the axion field remains stable around the horizon \cite{Branco:2023frw,Omiya:2022gwu}. Recent studies \cite{Takahashi:2021yhy, Takahashi:2021eso, Takahashi:2023flk} have described the behavior of axion clouds, focusing on how they evaporate and backreact on the spacetime during the merger of black hole binaries \cite{Yang:2017lpm}, a key way to look for signals of dark matter clouds \cite{Fukuda:2019ewf,Filippini:2019cqk,Banerjee:2019xds,Choudhary:2020pxy,Delgado:2020hwr,Zhang:2022rex,Bamber:2022pbs,Herdeiro:2023roz,Leong:2023nuk,Aurrekoetxea:2023jwk,Aurrekoetxea:2024cqd}. Beyond these clouds, the interaction between axion fields and gravity can lead to the formation of compact objects known as axion stars \cite{Guerra:2019srj,Delgado:2020udb,Zeng:2021oez,Zeng:2023hvq}. Axion fields have been extensively studied across multiple astrophysical contexts, motivated in particular by their intriguing couplings to electromagnetic fields \cite{Blas:2020nbs, Sakurai:2023hkg,Yoshino:2013ofa, Spieksma:2023vwl, Boskovic:2018lkj,Caputo:2024oqc}. References \cite{Arvanitaki:2016qwi, Arvanitaki:2014wva, Caputo:2025oap} show how we can use these effects to set limits on dark matter in future observations. Recently, new ideas such as squeezed gravitons from superradiant clouds \cite{Dorlis:2025zzz,Dorlis:2025amf,Mavromatos:2025ofn} have also gained attention, showing many ways axions and black holes can interact.

Considering the strong connection between black hole superradiance and axion fields, this paper investigates an Einstein-Maxwell system coupled with a charged axion field. In our model, the scalar field with an axion potential grows through superradiance, extracting both charge and energy from the black hole to eventually form axion clouds. By employing the spherical Baumgarte-Shibata-Shapiro-Nakamura formulation \cite{Alcubierre:2011pkc,Baumgarte:2012xy,Cordero-Carrion:2012qac,Montero:2012yr}, we obtain the detailed evolution of the axion field. Our results show how energy and charge are extracted, leading to the formation of black holes with scalar hair.

This paper is organized as follows. In Section \ref{sec:model}, we introduce the physical system and the formalism used to evolve the spacetime and matter sources. We also briefly describe the numerical methods employed to solve the equations of motion. In Section \ref{sec:results}, we present our main results, provide physical explanations, and compare our findings with previous research. Finally, Section \ref{sec:conc} concludes the paper with a summary and a discussion of future work. Throughout this work, we use geometric units where $G=c=\hbar=1.$

\section{Model} \label{sec:model}

To begin, we introduce a charged axion field $\phi$. In a spherical RNBH spacetime, the action takes the following form
\begin{eqnarray}
S=\int d^4x\sqrt{-g}\bigg(\frac{R-\mathcal{I}}{16\pi}-\frac{1}{2}g^{\mu\nu}D_{\mu}\phi^*D_{\nu}\phi-V(\phi)\bigg)
\label{action}
\end{eqnarray}
with
\begin{equation}
 \mathcal{I}=F_{\alpha\beta}F^{\alpha\beta}.
 \end{equation}
Here, the covariant derivation act as
\begin{equation}
D_{\mu}=\nabla_{\mu}+iqA_{\mu},
\end{equation}
where $A_{\mu}$ is $U(1)$ vector field and $q$ is the charge of an axion particle. The scalar potential $V(\phi)$ of axion is taken as \cite{Delgado:2020hwr,GrillidiCortona:2015jxo,Delgado:2020udb}
\begin{equation}
    \begin{split}
        V(\phi)=\frac{m^2f^2_a}{B}\left[1-\sqrt{1-4B\sin^2\left(\frac{|\phi|}{2f_a}\right)}\right].
        \label{axionpotential}
    \end{split}
\end{equation}
Here, $m$ and $f_a$ represent the mass and decay constant of the axion, respectively. The parameter $B \approx 0.22$ is defined by $B = z/(1+z)^2$, where $z \equiv m_u/m_d \approx 0.48$ denotes the mass ratio of the up and down quarks \cite{GrillidiCortona:2015jxo}. The axion potential can be expanded as follows
\begin{equation}
    \begin{split}
        \phi(V)= &\frac{1}{2}m^2|\phi|^2+\left(\frac{3B-1}{24}\right)\frac{m^2}{f_a^2}|\phi|^4\\
        &+\frac{1+15B(3B-1)}{720 f_a^4}m^2|\phi|^6\cdots.
    \end{split}
\end{equation}
From this expansion, it is clear that $m$ determines the mass of the field, while $f_a$ controls the strength of the quartic and higher-order self-interactions.

Varying the action \eqref{action} with respect to $g_{\mu\nu}$, $A_\mu$, and $\phi$, we can get the following equations of motion:
\begin{eqnarray}
        R_{\mu\nu}-\frac{1}{2}g_{\mu\nu}R=8\pi(T^{(\phi)}_{\mu\nu}+T^{(em)}_{\mu\nu}),\\
        \nabla^{\mu}F_{\mu\nu}=2\pi iq[\phi^*D_\nu\phi-\phi D_\nu\phi^*]:=4\pi j^{e\nu},\\
        \nabla_\mu\nabla^\mu\phi+iqA^\alpha(2\nabla_\alpha\phi+iqA_\alpha\phi)+iq\phi\nabla_\beta A^\beta=\frac{dV}{d\phi}.
\end{eqnarray}
The energy-momentum tensor of matter sources are as follows
\begin{eqnarray}
        T^{(em)}_{\mu\nu}  &=&\frac{1}{4\pi}\left(F_{\mu\alpha}F^{\alpha}_\nu-\frac{1}{4}g_{\mu\nu}F^{\alpha\beta}F_{\alpha\beta}\right),\label{TmunuEM}\\
        T^{(\phi)}_{\mu\nu} &=&\frac{1}{2}(D_\mu\phi^*)(D_\nu\phi)+\frac{1}{2}(D_\mu\phi)(D_\nu\phi^*)\nonumber\\
          &&-g_{\mu\nu}[(D^\alpha\phi^*)D_\alpha\phi+V(\phi)].\label{Tmunuaxion}
\end{eqnarray}
We follow the convention that $\phi$ is dimensionless and $\mu$ has dimensions of (length)$^{-1}$.

In the next part we show the BSSN formulations we use in our numerical evolution, including matter source and spacetime term \cite{Baumgarte:2012xy,Cordero-Carrion:2012qac,Sanchis-Gual:2016tcm}.
The lapse function $\alpha$ is evolved with the 1+log condition \cite{Bona:1997hp}
\begin{equation}
    \begin{split}
        \partial_t\alpha=-2\alpha K,
    \end{split}
\end{equation}
and the shift vector $\beta^r$ is evolved with a variation of the gamma-driver condition \cite{Alcubierre:2002kk}
\begin{equation}
    \begin{split}
        \partial_t\beta^r=B^r,~\partial_tB^r=\frac{3}{4}\partial_t\hat{\Delta}^r-2B^r.
    \end{split}
\end{equation}
On spacelike hypersurface we have two constraints namely Hamiltonian constraint
\begin{equation}
    \begin{split}
         R-(A_a^2+2A_b^2)+\frac{2}{3}K^2-16\pi \mathcal{E}=0,
    \end{split}
\end{equation}
and momentum constraint \cite{Baumgarte:2012xy}
\begin{equation}
        \partial_rA_a-\frac{2}{3}\partial_rK+6A_a\partial_r\chi +(A_a-A_b)\left(\frac{2}{r}+\frac{\partial_rb}{b}\right)=8\pi j_r.
\end{equation}
Here, $\mathcal{E}=\mathcal{E}^{(\phi)}+\mathcal{E}^\text{(em)}$ and $j_r=j^{(\phi)}_{r}+j^\text{(em)}_{r}$ denote the total energy and momentum densities, representing the combined contributions from the scalar and electromagnetic fields. Their explicit definitions will be provided in the subsequent sections.

We now derive the evolution equations for the matter fields from energy-momentum tensors \eqref{TmunuEM} and \eqref{Tmunuaxion}. The spacetime is described by the $3+1$ metric split
\begin{eqnarray}
ds^2 = -\alpha^2 dt^2 + \gamma_{ij} (dx^i + \beta^i dt)(dx^j + \beta^j dt),
\end{eqnarray}
which, under spherical symmetry, reduces to
\begin{equation}
ds^2 = (-\alpha^2 + \beta_r \beta^r) dt^2 + 2\beta_r dt dr + e^{4\chi} (a dr^2 + r^2 b d\Omega^2).
\end{equation}
Here, $\alpha, \beta^r, a, b,$ and $\chi$ are functions of $(t, r)$, and $d\Omega^2=d\theta^2+\sin^2\theta d\varphi^2$ denotes the standard metric on the 2-sphere. The induced metric $\gamma_{ij}$ on the spacelike hypersurface is
\begin{equation}
    \begin{split}
    \gamma_{ij}=e^{4\chi}\text{diag}(a,br^2,br^2sin^2\theta).
    \end{split}
\end{equation}
We perform the numerical simulations by using our spherical numerical relativity code, see Refs. \cite{Zhang:2023qxf,Zhang:2024wci} for convergence analysis and technical details.

\subsection{Field Variables and Physical Observables}

Following the formalism in \cite{Alcubierre:2009ij}, we decompose the vector potential $A^\mu$ by introducing the scalar potential $\Phi$ and the spatial vector potential $a^i$ as measured by an Eulerian observer
\begin{equation}
\Phi = -n_\mu A^\mu, \quad a^i = \gamma^i_{\, \mu} A^\mu.
\end{equation}
Under the assumption of spherical symmetry, the electric and magnetic fields measured by these Eulerian observers are defined as
\begin{equation}
E^{\mu} = F^{\mu\nu} n_{\nu}, \quad B^{\mu} = {}^*F^{\mu\nu} n_{\nu},
\end{equation}
where ${}^*F^{\mu\nu} = \frac{1}{2} \epsilon^{\mu\nu\alpha\beta} F_{\alpha\beta}$ is the dual field strength tensor. We adopt the convention for the Levi-Civita tensor such that $\epsilon^{0123} = -1/\sqrt{-g}$ and $\epsilon_{0123} = \sqrt{-g}$, where $g$ denotes the determinant of the four-dimensional metric.

Here, $F^{*\mu\nu}=\frac{1}{2}\epsilon^{\mu\nu\alpha\beta}F_{\alpha\beta}.$ We take the convention that $
\epsilon^{0123}=\frac{-1}{\sqrt{-g}}$ and $\epsilon_{0123}=+\sqrt{-g}$, where g is the determinant of the four-dimensional metric.

For the electromagnetic sector, spherical symmetry implies that the electric field $E^\mu$ possesses only a radial component, while the magnetic field $B^\mu$ vanishes identically. Following the formulations in \cite{Torres:2014fga,Corelli:2021ikv}, the evolution equations for the electromagnetic potentials and the radial electric field are given by
\begin{equation}
\begin{split}
\partial_t \Phi = & \beta^r \partial_r \Phi + \alpha K \Phi \\
& - \frac{\alpha}{a e^{4\chi}} \left[ \partial_r a_r + a_r \left( \frac{2}{r} - \frac{\partial_r a}{2a} + \frac{\partial_r b}{b} + 2 \partial_r \chi \right) \right] \\
& - \frac{\alpha}{a e^{4\chi}} \partial_r \alpha,
\end{split}
\end{equation}
\begin{equation}
\partial_t a_r = \beta^r \partial_r a_r + a_r \partial_r \beta^r - a \alpha e^{4\chi} E^r - \partial_r (\Phi \alpha),
\end{equation}
\begin{equation}
\partial_t E^r = \beta^r \partial_r E^r - E^r \partial_r \beta^r + \alpha K E^r - 4\pi \alpha j^{er},
\end{equation}
where $K$ denotes the trace of the extrinsic curvature $K_{ij}$, and $j^{er}$ is the electric current density as measured by the Eulerian observers.

To solve the Klein-Gordon equation numerically, we recast it into a first-order system by introducing two auxiliary variables: the conjugate momentum $\Pi$ and the spatial gradient $\Psi$, defined as
\begin{equation}
\Pi := n^\alpha \partial_\alpha \phi = \frac{1}{\alpha} (\partial_t \phi - \beta^r \partial_r \phi),
\end{equation}
\begin{equation}
\Psi := \partial_r \phi.
\end{equation}
Consequently, the second-order Klein-Gordon equation is transformed into the following set of first-order evolution equations:
\begin{equation}
\partial_t \phi = \beta^r \partial_r \phi + \alpha \Pi,
\end{equation}
\begin{equation}
\partial_t \Psi = \beta^r \partial_r \Psi + \Psi \partial_r \beta^r + \partial_r (\alpha \Pi),
\end{equation}
\begin{widetext}
\begin{equation}
\begin{split}
\partial_t \Pi = & \beta^r \partial_r \Pi + \alpha K \Pi + \frac{\alpha}{a e^{4\chi}} \left[ \partial_r \Psi + \Psi \left( \frac{2}{r} - \frac{\partial_r a}{2a} + \frac{\partial_r b}{b} + 2 \partial_r \chi \right) \right] - \alpha \left[ \frac{dV}{d\phi} + q^2 \left( \frac{a_r^2}{a e^{4\chi}} - \Phi^2 \right) \right] \phi \\
& + \frac{\Psi}{a e^{4\chi}} \partial_r \alpha+ 2i q \alpha \left( \frac{a_r \Psi}{a e^{4\chi}} + \Phi \Pi \right),
\end{split}
\end{equation}
\end{widetext}
where the derivative of the scalar potential $V(\phi)$ with respect to the field is given by
\begin{equation}
\frac{dV}{d\phi} = \frac{\mu^2 f_a \phi_R \sin \left( \sqrt{\phi_R^2 + \phi_I^2} / f_a \right)}{\sqrt{\phi_R^2 + \phi_I^2} \sqrt{1 - 2B + 2B \cos \left( \sqrt{\phi_I^2 + \phi_R^2} / f_a \right)}}.
\end{equation}

Following Ref. \cite{Sanchis-Gual:2016tcm}, we define the gauge-invariant versions of the variables $\Pi$ and $\Psi$.
\begin{equation}
    \begin{split}
        \tilde{\Pi}= n^{\mu}D_{\mu}\phi^*=\Pi-iq\Phi\phi,
    \end{split}
\end{equation}
\begin{equation}
    \begin{split}
        \tilde{\Psi}= \gamma^{\mu}_rD_{\mu}\phi=\Psi+iqa_r\phi.
    \end{split}
\end{equation}
Consequently, the matter source terms arising from the scalar field's energy-momentum tensor take the form
\begin{eqnarray}
  \mathcal{E}^{(\phi)}&=& n^{\mu}n^{\nu}T^{(\phi)}_{\mu\nu}=\frac{1}{2}\left(\tilde{|\Pi|}^2+\frac{|\tilde{\Psi}|^2}{ae^{4\chi}}\right)+V(\phi),\\
  j^{(\phi)}_{r}&=&-\gamma^{\alpha}_rn^\beta T^{(\phi)}_{\alpha\beta}=-\frac{1}{2}\left(\tilde{\Pi}^*\tilde{\Psi}+\tilde{\Psi}^*\tilde{\Pi}\right),
\end{eqnarray}

\begin{eqnarray}
        S^{(\phi)}_a&=& (T^r_r)^{(\phi)}=\frac{1}{2}\left(\tilde{|\Pi|}^2+\frac{|\tilde{\Psi}|^2}{ae^{4\chi}}\right)-V(\phi),\\
        S^{(\phi)}_b&=& (T^r_r)^{(\phi)}=\frac{1}{2}\left(\tilde{|\Pi|}^2-\frac{|\tilde{\Psi}|^2}{ae^{4\chi}}\right)-V(\phi),
\end{eqnarray}
and for the electric field
\begin{eqnarray}
        \mathcal{E}^\text{(em)}&=&\frac{1}{8\pi}ae^{4\chi}(E^r)^2,\\
        S^\text{(em)}_a&=&-\frac{1}{8\pi}ae^{4\chi}(E^r)^2,\label{Q_density_sf}\\
        S^\text{(em)}_b&=&\frac{1}{8\pi}ae^{4\chi}(E^r)^2.
\end{eqnarray}
The electromagnetic momentum density $j^\text{(em)}_r$ vanishes as the magnetic field is zero under spherical symmetry. Integrating the energy density yields the total energy of the axion field
\begin{equation}
    \begin{split}
        E^{(\phi)}=\int^{r_m}_{r_{AH}}\mathcal{E}^{(\phi)}dV.
    \end{split}
\end{equation}
Furthermore, we calculate the charge of the black hole according to the prescription in \cite{Sanchis-Gual:2016tcm} at the apparent horizon
\begin{equation}
    \begin{split}
        Q_\text{BH}=(r^2e^{6\chi}\sqrt{ab^2}E^r)|_\text{AH}.
    \end{split}
\end{equation}
The charge of scalar field is obtained by substituting the following charge density $\rho_q$ for $\mathcal{E}^{(\phi)}$
\begin{equation}
\rho_q = q(\Pi_I\phi_R-\Pi_r\phi_I)-q^2\Phi|\phi_I^2+\phi_R^2|.
\end{equation}

\subsection{Initial data}

In our model of a cavity-enclosed RNBH, any small-amplitude scalar field is sufficient to trigger superradiant instability, regardless of its specific spatial profile. This instability allows for the extraction of energy and charge from the event horizon. Therefore, we set the initial configuration as an RNBH perturbed by a scalar field with a radial Gaussian distribution as follows
\begin{equation}
    \begin{split}
        \phi=A_0 e^{-\frac{(r-r_0)^2}{\sigma^2}}.
    \end{split}
\end{equation}
The parameters $A_0, r_0,$ and $\sigma$ characterize the initial Gaussian distribution.

To ensure the scalar field remains confined, we require the amplitude to vanish at the mirror boundary, $r = r_{\text{mirr}}$, leading to the following boundary conditions
\begin{equation}
    \begin{split}
        &\phi(r_{\text{mirr}})=\Psi(r_{\text{mirr}})=\Pi(r_{\text{mirr}})=0,\\
        &\partial_r\phi(r_{\text{mirr}})=\partial_r\Psi(r_{\text{mirr}})=\partial_r\Pi(r_{\text{mirr}})=0.
    \end{split}
\end{equation}
The associated auxiliary first-order quantities are initialized as follows
\begin{equation}
    \begin{split}
        \Pi(t=0,r)=0.
    \end{split}
\end{equation}

Since the initial scalar field amplitude is infinitesimal, its backreaction on the spacetime geometry at $t=0$ is negligible. Consequently, the initial background is described entirely by the RN metric. To facilitate numerical evolution, we adopt isotropic coordinates, in which the spatial 3-metric is conformally flat (i.e., $a=b=1$). Furthermore, we impose the time-symmetry condition, setting the extrinsic curvature to $K_{ij}=0$. In this coordinate system, the spatial 3-metric is expressed as
\begin{equation}
    \begin{split}
        dl^2=\psi^4(dr^2+r^2d\Omega^2),
    \end{split}
\end{equation}
where the conformal factor $\psi$ is given by
\begin{equation}
\psi = \left[ \left(1 + \frac{M}{2r} \right)^2 - \frac{Q^2}{4r^2} \right]^{1/2}.
\end{equation}
Here, $M$ and $Q$ are the black hole mass and total charge, respectively. At $t=0$, we adopt a ``pre-collapsed" lapse profile, $\alpha = \psi^{-2}$, and a vanishing shift vector, $\beta^r = 0$. The initial radial electric field is then specified as
\begin{equation}
E^r = \frac{Q}{r^2 \psi^6}.
\end{equation}

Having established the numerical framework, we now proceed to a comprehensive parametric study of the system's evolution. Our analysis focuses on how the core parameters ($Q, q, f_a, m$) dictate the behavior of the observables $E$ and $\phi$. Specifically, we evolve the system in time to elucidate the axion field dynamics and the resulting charge exchange mechanisms. This allows for a detailed comparison of the superradiance rate and extraction efficiency across different physical regimes.

For all simulations, we set $M=1$ to fix the energy scale of the system. The dimensionless decay constant is defined as $\bar{f_a} = f_a / \phi_0$, where $\phi_0$ is a constant with the dimension of the scalar field, which we set to unity. We explore the following values for the decay constant:
\begin{equation}
\bar{f_a} \in \{0.0001, 0.001, 0.01, 0.1, 1, 20\}.
\end{equation}
To investigate the influence of the cavity size, we evolve the system with mirror radii $r_{\text{mirr}}/M \in \{10, 15\}$. Given the charged nature of the background, we characterize the charge transfer by considering a range of axion field charges
\begin{equation}
\bar{q} = qM \in \{5, 15, 20, 50\}.
\end{equation}
The parameters governing the initial Gaussian pulse are fixed at $A_0 = 0.0001, r_0 = 5,$ and $\sigma = 1$. Finally, the scalar mass parameter is varied among the following choices
\begin{equation}
\bar{m} = mM \in \{0, 0.2, 0.4, 0.6\}.
\end{equation}
Throughout our work, we extract field strength at $r_\text{extract} = 5M$.

\section{Results} \label{sec:results}

\begin{figure}[htbp]
	\begin{center}
		\includegraphics[width=0.9\linewidth]{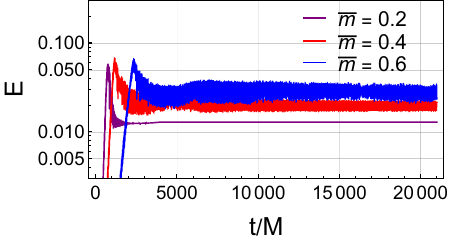}
	\end{center}
	\caption{Energy evolution of the axion field for various mass parameters within a cavity of radius $r_{\text{mirr}}=20M$. The parameters are set to $\bar{f_a}\to\infty$ and $\bar{q}=20$.}
	\label{Ct}
\end{figure}

Before presenting our primary findings, we address the parameter selection and the robustness of our numerical framework. To rigorously verify the accuracy of our implementation, we first reproduced the results of previously published studies in the literature \cite{Sanchis-Gual:2015lje,Sanchis-Gual:2016tcm}. Beyond the reproduction of existing work, we conducted comprehensive stability tests to ensure the long-term reliability of our simulations, as illustrated in Fig. \ref{Ct}. Our findings show that, depending on the chosen parameter sets, the system eventually settles into either a stationary state or a stable oscillatory regime characterized by a rapid and constant frequency. Our code yielded results in excellent agreement with established benchmarks, confirming that our numerical solver correctly captures the underlying physics.

Furthermore, as shown in Figs. \ref{mu0.2mirr10grp1E} and \ref{mu0.2mirr10grp1trans}, we initially examined six values of the decay constant $\bar{f_a}$. These results demonstrate that increasing $\bar{f_a}$ beyond a certain threshold does not significantly alter the system's dynamics. Consequently, to optimize computational efficiency without loss of physical generality, we restrict our subsequent analysis to four representative values of $\bar{f_a}$.

These tests, combined with our successful reproduction of prior literature, verify that our code is well-equipped to handle long-term evolution while maintaining high numerical precision. Having validated our solver, we now proceed to analyze the results. The data presented hereafter constitute a representative subset of our simulations, carefully selected to highlight the typical trends and the most salient physical phenomena in the axion-black hole system.

\subsection{The influence of $f_a$ and $q$}

To evaluate the dynamical properties of the system, we adopt a controlled-variable approach across the parameter space $\{\bar{f_a}, \bar{q}, \bar{m}, r_{\text{mirr}}\}$. Notably, as the axion potential \eqref{axionpotential} vanishes in the massless limit ($\bar{m}=0$), rendering the decay constant $\bar{f_a}$ irrelevant, we perform our simulations with a fixed non-zero mass $\bar{m}$ to analyze the effects of $\bar{f_a}$ and $\bar{q}$. We consider two cavity radii, $r_{\text{mirr}}=10M$ and $r_{\text{mirr}}=15M$, to investigate how boundary proximity influences the energy extraction process.

\begin{figure*}[htbp]
	\begin{center}
		\includegraphics[width=0.3\linewidth]{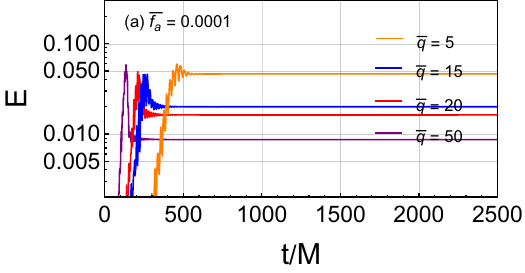}
        \includegraphics[width=0.3\linewidth]{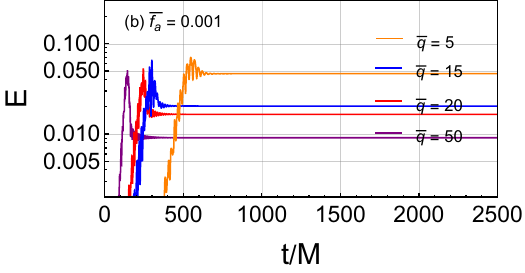}
        \includegraphics[width=0.3\linewidth]{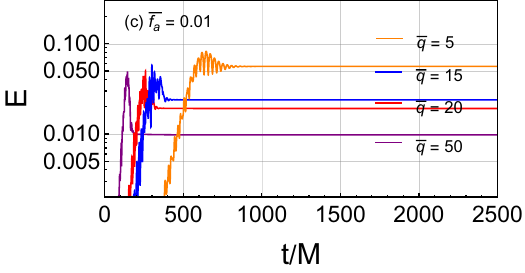}

        \includegraphics[width=0.3\linewidth]{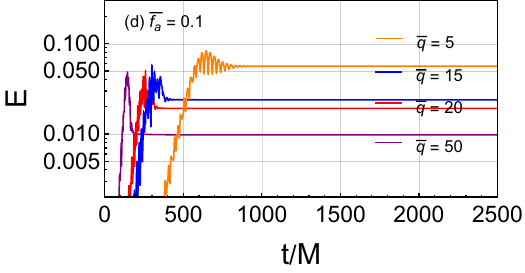}
        \includegraphics[width=0.3\linewidth]{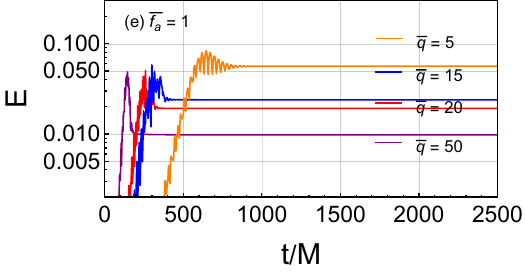}
        \includegraphics[width=0.3\linewidth]{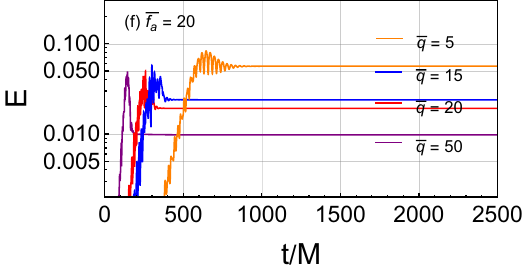}
	\end{center}
	\caption{Energy evolution of the axion field for various decay constants $\bar{f_a}$ and $\bar{q}$ within a cavity of radius $r_{\text{mirr}}=10M$. The parameter is set to $\bar{m}=0.2$.}
	\label{mu0.2mirr10grp1E}
\end{figure*}

\begin{figure*}[htbp]
	\begin{center}
		\includegraphics[width=0.3\linewidth]{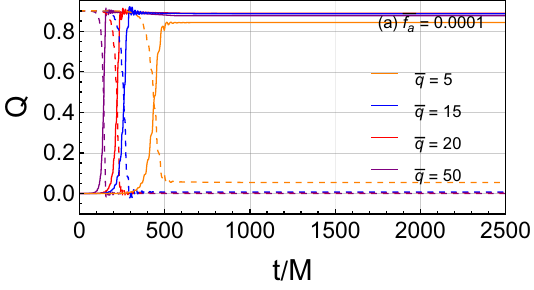}
        \includegraphics[width=0.3\linewidth]{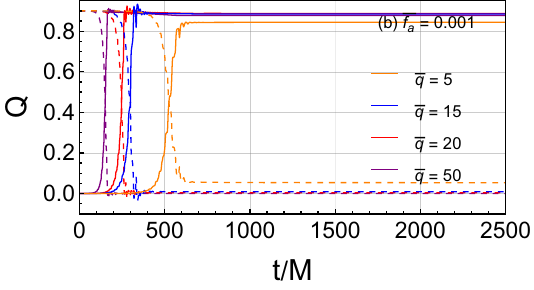}
        \includegraphics[width=0.3\linewidth]{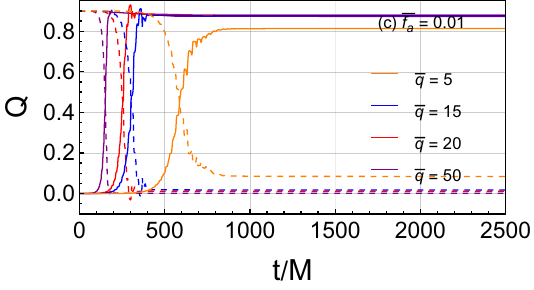}

        \includegraphics[width=0.3\linewidth]{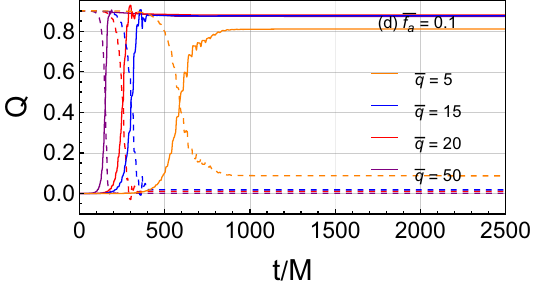}
        \includegraphics[width=0.3\linewidth]{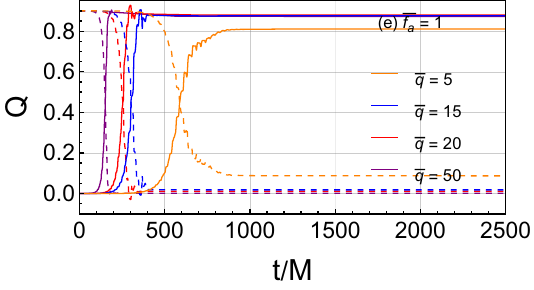}
        \includegraphics[width=0.3\linewidth]{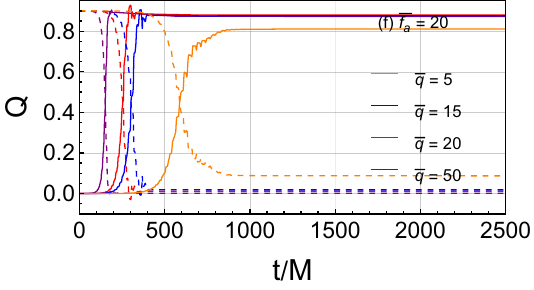}
	\end{center}
	\caption{Charge transfer between the axion field and the black hole for various decay constants $\bar{f_a}$ and $\bar{q}$ within a cavity of radius $r_{\text{mirr}}=10M$. Dashed and solid lines denote the charges of the black hole and the axion field, respectively. The parameter is set to $\bar{m}=0.2$.}
	\label{mu0.2mirr10grp1trans}
\end{figure*}

In the more compact cavity (Figs. \ref{mu0.2mirr10grp1E} and \ref{mu0.2mirr10grp1trans}) with $r_\text{mirr}=10M$, an increase in the field charge $\bar{q}$ leads to a reduction in total energy and enhancement for charge transfer but an acceleration in the superradiance rate, consistent with Ref. \cite{Sanchis-Gual:2016tcm}. The influence of the decay constant $\bar{f_a}$ is most pronounced at small values and gradually diminishes as it increases. The system exhibits a clear sensitivity to $\bar{f_a}$ at lower values, however, this effect tapers off steadily. Upon reaching the scale of $0.01$, the influence of $\bar{f_a}$ on the final energy and dynamical evolution becomes practically negligible.

\begin{figure*}[htbp]
	\begin{center}
		\includegraphics[width=0.43\linewidth]{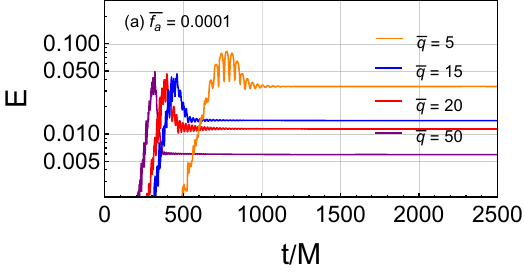}
        \includegraphics[width=0.43\linewidth]{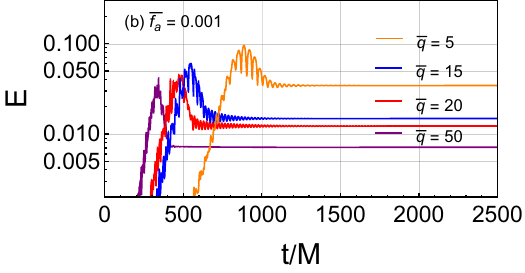}
        \includegraphics[width=0.43\linewidth]{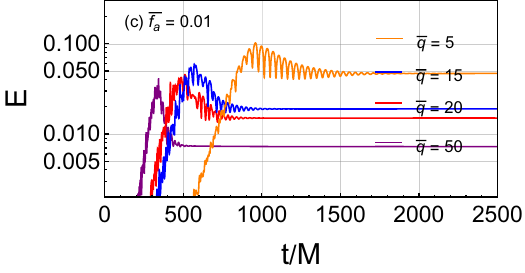}
        \includegraphics[width=0.43\linewidth]{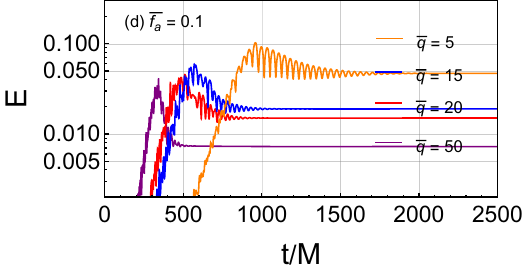}
	\end{center}
	\caption{Energy evolution of the axion field for various decay constants $\bar{f_a}$ and $\bar{q}$ within a cavity of radius $r_{\text{mirr}}=15M$. The parameter is set to $\bar{m}=0.2$.}
	\label{mu0.2mirr15grp1E}
\end{figure*}

\begin{figure*}[!htbp]
	\begin{center}
		\includegraphics[width=0.43\linewidth]{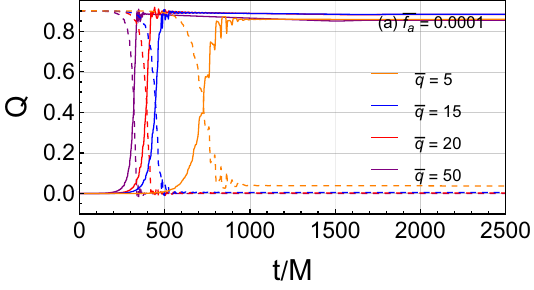}
        \includegraphics[width=0.43\linewidth]{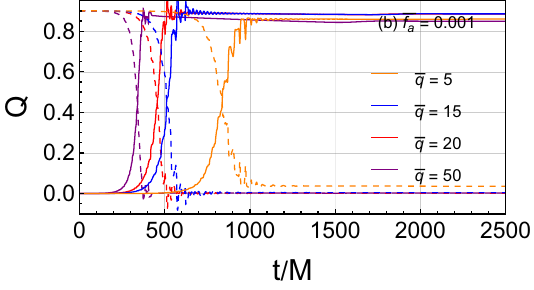}
        \includegraphics[width=0.43\linewidth]{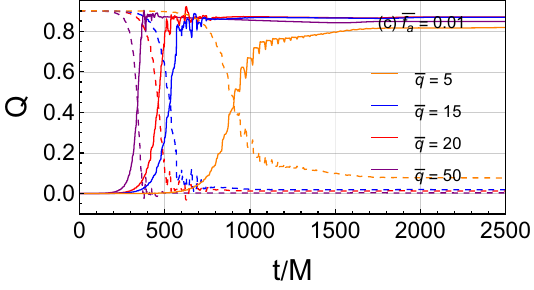}
        \includegraphics[width=0.43\linewidth]{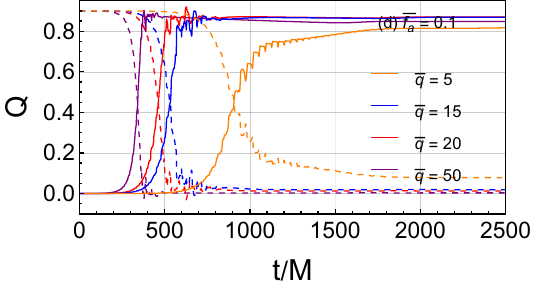}
	\end{center}
	\caption{Charge transfer between the axion field and the black hole for various decay constants $\bar{f_a}$ and $\bar{q}$ within a cavity of radius $r_{\text{mirr}}=15M$. Dashed and solid lines denote the charges of the black hole and the axion field, respectively. The parameter is set to $\bar{m}=0.2$.}
	\label{mu0.2mirr15grp1trans}
\end{figure*}

In contrast, for the larger cavity radius with $r_{\text{mirr}}=15M$ (Figs. \ref{mu0.2mirr15grp1E} and \ref{mu0.2mirr15grp1trans}), the overall energy extraction and charge transition are weaker and the evolution proceeds more slowly. Interestingly, in this regime, the impact of the field charge $\bar{q}$ becomes more pronounced, particularly in the higher decay region (as $\bar{f_a}$ reaches the scale of $0.01$). Despite these quantitative differences, the qualitative trends align with the $10M$ case.

A comparison between these two scenarios reveals that the proximity of the mirror in the $r_{\text{mirr}}=10M$ case tends to suppress the influence of other parameters. This suggests that bringing the unphysical boundary closer to the black hole accelerates charge extraction at the expense of energy extraction efficiency. We hypothesize that when one parameter (such as $r_{\text{mirr}}$) is set to an extreme value, it can dominate the system's response, thereby masking the effects of secondary parameters. Consequently, a comprehensive understanding of superradiance requires a careful balancing of criticality and boundary effects.

\subsection{The influence of $m$}

\begin{figure*}[htbp]
	\begin{center}
		\includegraphics[width=0.43\linewidth]{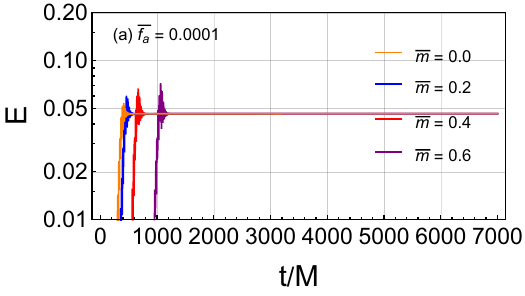}
        \includegraphics[width=0.43\linewidth]{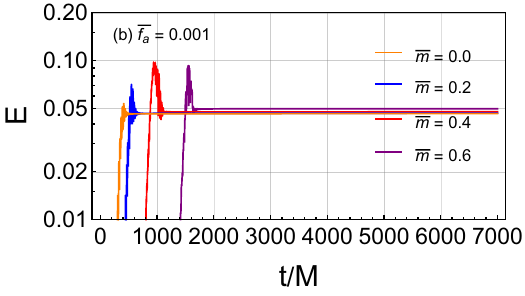}
        \includegraphics[width=0.43\linewidth]{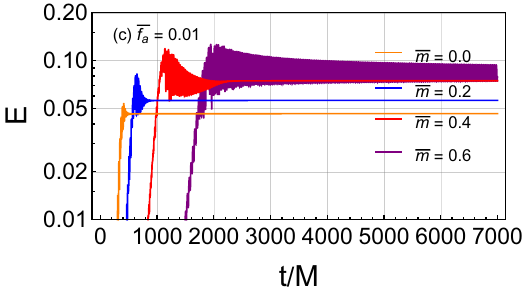}
        \includegraphics[width=0.43\linewidth]{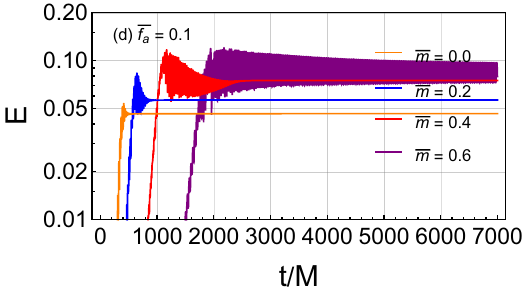}
	\end{center}
	\caption{Energy evolution of the axion field for various decay constants $\bar{f_a}$ and mass parameters within a cavity of radius $r_{\text{mirr}}=10M$. The parameter is set to $\bar{q}=5$.}
	\label{mirr10grp4}
\end{figure*}

\begin{figure*}[htbp]
	\begin{center}
		\includegraphics[width=0.43\linewidth]{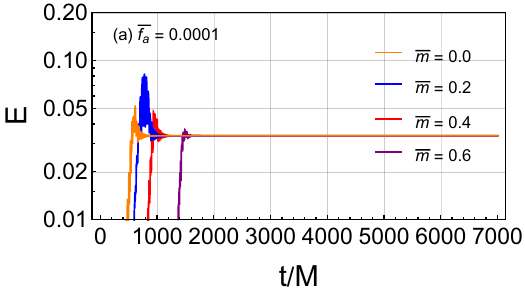}
        \includegraphics[width=0.43\linewidth]{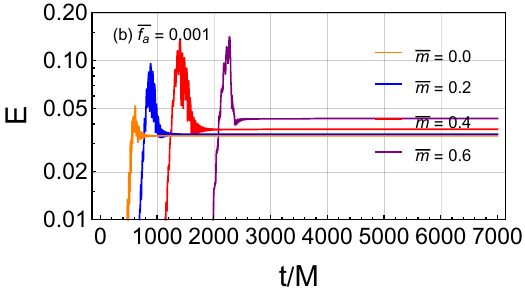}
        \includegraphics[width=0.43\linewidth]{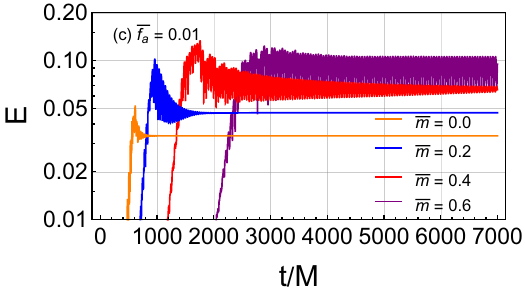}
        \includegraphics[width=0.43\linewidth]{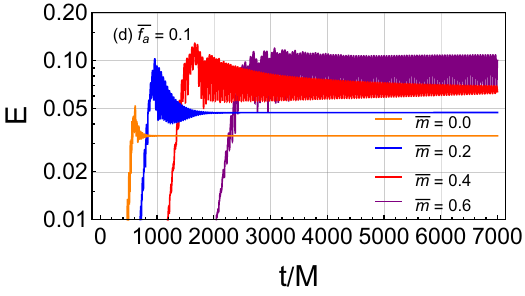}
	\end{center}
	\caption{Energy evolution of the axion field for various decay constants $\bar{f_a}$ and mass parameters within a cavity of radius $r_{\text{mirr}}=15M$. The parameter is set to $\bar{q}=5$.}
	\label{mirr15grp4}
\end{figure*}

We now shift our focus to the scalar mass parameter $\bar{m}$, whose impact on the system is considerably more pronounced than that of the decay constant or field charge. The dependence of the energy extraction process on $\bar{m}$ is illustrated in Figs. \ref{mirr10grp4} and \ref{mirr15grp4} for fixed cavity radii $r_{\text{mirr}}=10M$ and $r_{\text{mirr}}=15M$.


Our results reveal that the role of the mass parameter is highly sensitive to the magnitude of the decay constant $\bar{f_a}$: in the low decay constant ($\bar{f_a} \leq 0.001$) regime, an increase in the mass parameter $\bar{m}$ primarily decelerates the energy growth rate without significantly altering the final saturated energy level; in the high decay constant regime, at higher decay constants $\bar{f_a}$, adopting larger values of $\bar{m}$ not only slows the dynamical evolution but also leads to a systematic increase in the final energy acquired by the axion field.

This behavior can be understood through the structure of the axion potential. A lower decay constant $\bar{f_a}$ triggers stronger non-linear modes and a more robust superradiant instability. Since the mass parameter essentially acts as a scaling factor for the potential without altering its qualitative functional form. Consequently, while $\bar{m}$ governs the timescale of the growth, the non-linearity dictated by $\bar{f_a}$ determines the saturation point and the overall strength of the instability.

\begin{figure}[]
	\begin{center}
		\includegraphics[width=1\linewidth]{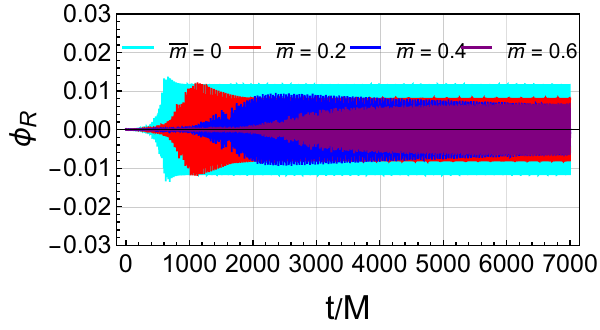}
        \includegraphics[width=1\linewidth]{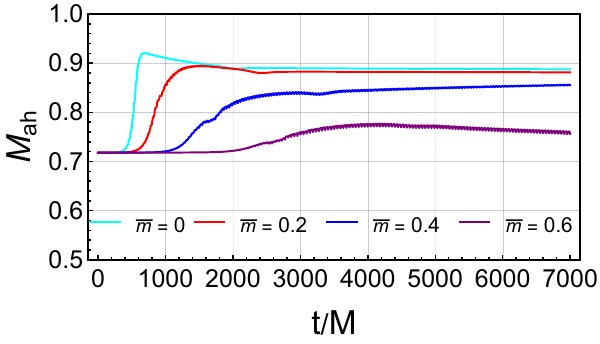}
	\end{center}
	\caption{Waveform of the axion field and irreducible mass of the black hole for various mass parameters within a cavity of radius $r_{\text{mirr}}=15M$. The parameters are set to $\bar{f_a}=0.1$ and $\bar{q}=5$.}
	\label{waveform}
\end{figure}

\begin{figure}[!htbp]
	\begin{center}
		\includegraphics[width=1\linewidth]{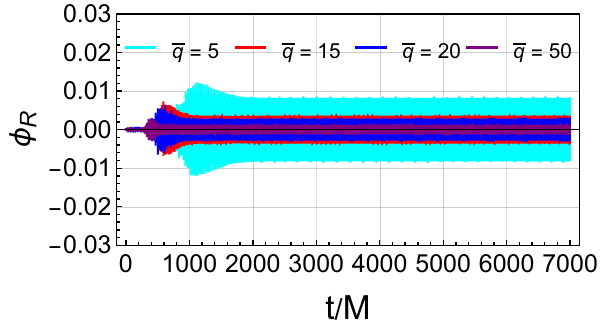}
        \includegraphics[width=1\linewidth]{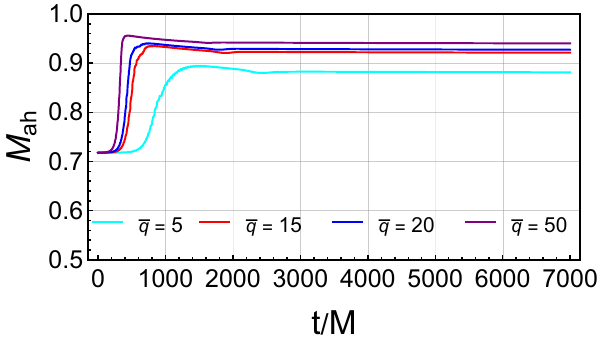}
	\end{center}
	\caption{Waveform of the axion field and irreducible mass of the black hole for various field charges $\bar{q}$ within a cavity of radius $r_{\text{mirr}}=15M$. The parameters are set to $\bar{f_a}=0.1$ and $\bar{m}=0.2$.}
	\label{waveformq}
\end{figure}

Beyond its impact on energy magnitude, the mass parameter $\bar{m}$ fundamentally alters the long-term dynamical evolution of the system when the value of decay constant increases. As illustrated in Fig. \ref{mirr15grp4}, we observe a distinct transition in the oscillatory behavior of the axion field. For $\bar{m}=0.6$, the system settles into a sustained, long-term oscillatory mode. In contrast, for $\bar{m}=0.4$, a different dynamical signature emerges: a transient oscillation whose frequency gradually shifts and attenuates.

Our analysis further reveals a striking parallel between the mirror position and the mass parameter. Increasing the cavity radius $r_{\text{mirr}}$ more readily induces these long-term oscillation modes, an effect qualitatively similar to increasing the mass $\bar{m}$. Furthermore, an expanded cavity leads to a marginal decrease in the final energy and a reduction in the superradiance rate. These diverse behaviors are explicitly contrasted in Fig. \ref{waveform}, where the distinct profiles of decaying modes versus slow-growth oscillatory modes are clearly visualized.

Apart from the primary influence of the mass parameter, our results indicate that the system's sensitivity to the decay constant $\bar{f_a}$ is regime-dependent. In the low-decay regime ($\bar{f_a} \leq 0.001$) and with a sufficiently close mirror, $\bar{m}$ exerts no appreciable effect on the final energy, only modulating the superradiance timescale and the peak energy. However, a transition occurs in the high-decay regime ($\bar{f_a} \geq 0.01$). In this domain, while the mass parameter remains a dominant factor in determining the dynamics, further increases in the decay constant yield negligible changes to the evolution. This suggests a saturation effect: mass-dominated regime: where $\bar{m}$ dictates the oscillation frequency and growth timescale, and the non-linear dynamics of the axion field become effectively insensitive to the specific value of $\bar{f_a}$ once the decay constant becomes sufficiently large. This finding aligns with our previous work \cite{Zhang:2024wci}, reinforcing that the impact of the mass parameter $\bar{m}$ is modulated by the magnitude of $\bar{f_a}$.

\begin{figure*}[htbp]
	\begin{center}
		\includegraphics[width=0.43\linewidth]{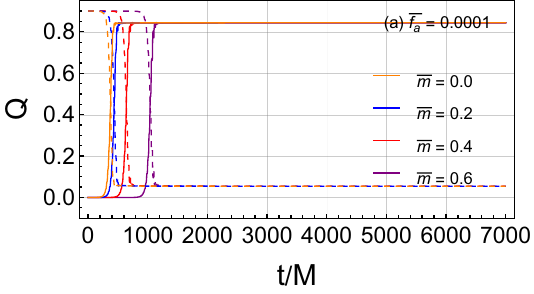}
        \includegraphics[width=0.43\linewidth]{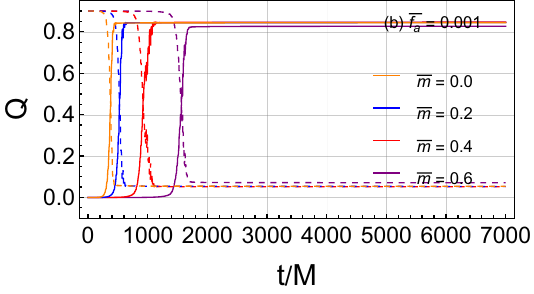}
        \includegraphics[width=0.43\linewidth]{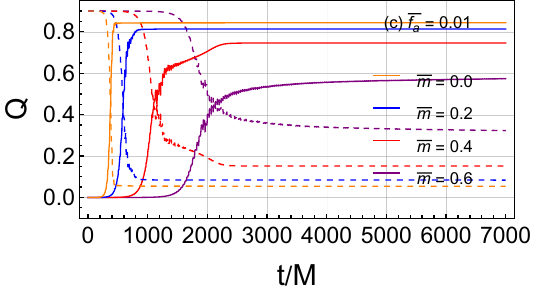}
        \includegraphics[width=0.43\linewidth]{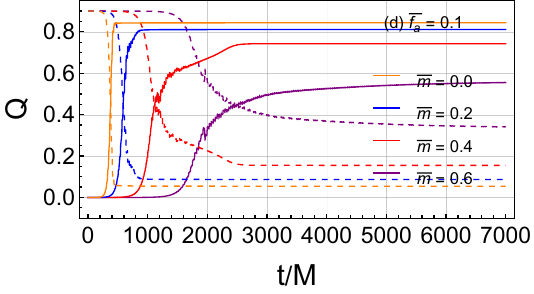}
	\end{center}
	\caption{Charge transfer between the axion field and the black hole for various decay constants $\bar{f_a}$ and mass parameters within a cavity of radius $r_{\text{mirr}}=10M$. Dashed and solid lines denote the charges of the black hole and the axion field, respectively. The parameter is set to $\bar{q}=5$.}
	\label{mirr10grp4trans}
\end{figure*}
\begin{figure*}[htbp]
	\begin{center}
		\includegraphics[width=0.43\linewidth]{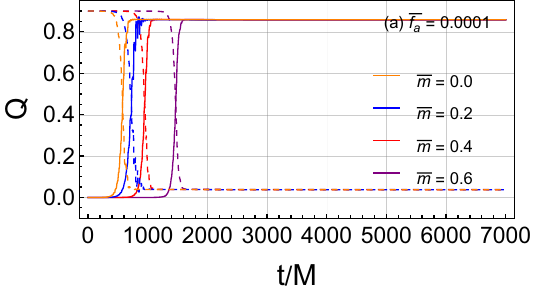}
        \includegraphics[width=0.43\linewidth]{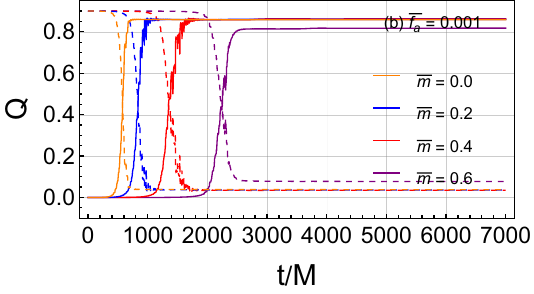}
        \includegraphics[width=0.43\linewidth]{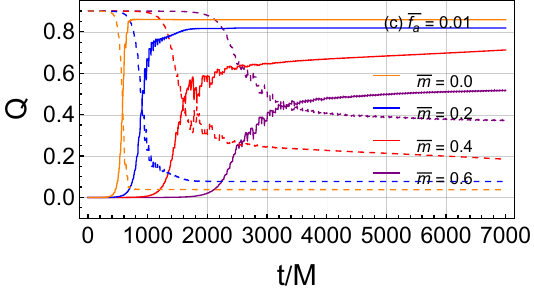}
        \includegraphics[width=0.43\linewidth]{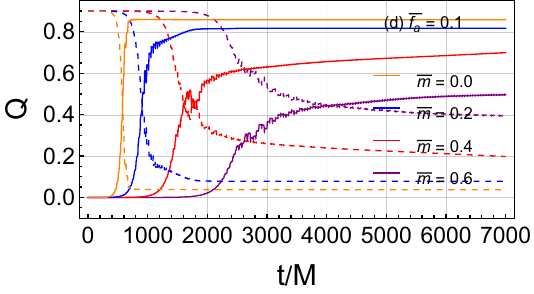}
	\end{center}
	\caption{Charge transfer between the axion field and the black hole for various decay constants $\bar{f_a}$ and mass parameters within a cavity of radius $r_{\text{mirr}}=15M$. Dashed and solid lines denote the charges of the black hole and the axion field, respectively. The parameter is set to $\bar{q}=5$.}
	\label{mirr15grp4trans}
\end{figure*}

Figures \ref{mirr10grp4trans} and \ref{mirr15grp4trans} illustrate the charge transfer dynamics for the aforementioned parameter sets. We observe that an increase in the mass parameter $\bar{m}$ not only extends the superradiance timescale but also leads to a more pronounced reduction in the final transferred charge. In the low-decay-constant regime, the final charge remains largely insensitive to $\bar{m}$, although the evolution is notably decelerated. Conversely, for larger $\bar{f_a}$, the charge transfer process is similarly delayed, but the final charge decreases significantly as $\bar{m}$ increases. This distinct sensitivity to the mass parameter within specific regions of the parameter space highlights a clear transition in the system's final state.

This behavior can be attributed to the interplay between non-linear modes and the decay constant. A sufficiently low $\bar{f_a}$ likely constrains the initial non-linear modes in high frequency regime, preventing the mass parameter from dominating the long-term evolution. In the high-mass regime, where energy extraction occurs without the characteristic 'beating' phenomenon, we posit that the prolonged superradiance timescale allows the fundamental modes to accumulate energy more efficiently, while higher-order modes fail to stabilize. These unstable modes eventually vanish because they are re-absorbed by the black hole; their frequencies fall outside the superradiant regime as the potential of the black hole decreases, a mechanism consistent with the findings in Ref. \cite{Sanchis-Gual:2016tcm}.

\subsection{properties of state at late time}

\begin{figure*}[htbp]
	\begin{center}
        \includegraphics[width=0.45\linewidth]{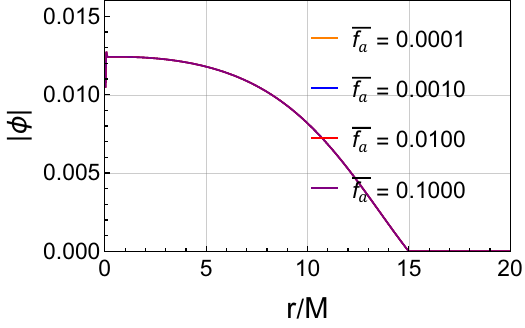}
        \includegraphics[width=0.45\linewidth]{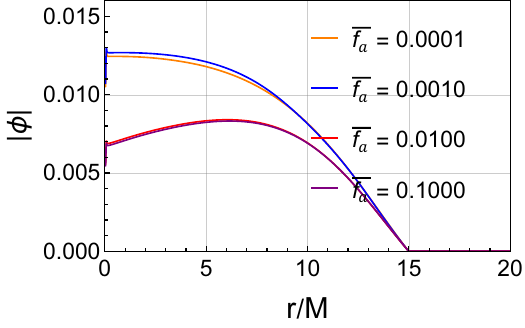}
	\end{center}
	\caption{Radial distribution of the axion field for two mass parameters $\bar{m} = \{0,0.2\}$, in left and right panel respectively, within a cavity of radius $r_{\text{mirr}}=15M$. The parameter is set to $\bar{q}=5$.}
	\label{distrismall1}
\end{figure*}

\begin{figure*}[htbp]
	\begin{center}
		\includegraphics[width=0.45\linewidth]{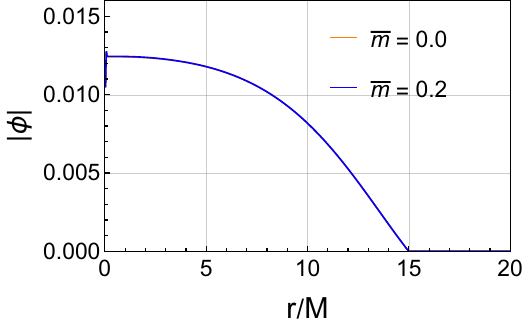}
        \includegraphics[width=0.45\linewidth]{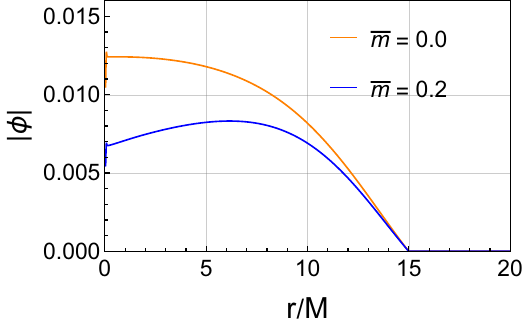}
	\end{center}
	\caption{Radial distribution of the axion field for two decay constants $\bar{f_a} = \{0.0001,0.1\}$, in left and right panel respectively, within a cavity of radius $r_{\text{mirr}}=15M$. The parameter is set to $\bar{q}=5$.}
	\label{distrismall2}
\end{figure*}

To further characterize the system's final state, we examine the influence of the mass parameter $\bar{m}$ and the decay constant $\bar{f_a}$ on the radial distribution of the axion field. Figures \ref{distrismall1} and \ref{distrismall2} illustrates the spatial morphology of the field as these parameters vary. In the low-decay-constant regime ($\bar{f_a}=0.0001$), the radial profile remains largely insensitive to the specific value of the mass parameter. Conversely, for $\bar{f_a}=0.1$, an increase in mass leads to a suppression of the total field strength, a trend consistent with our earlier energy analysis and the findings in Ref. \cite{Sanchis-Gual:2016tcm}. These results demonstrate that a stable, well-defined axion cloud forms within the cavity. The emergence of a distinct condensate region in the more massive cases underscores the pivotal role of the mass parameter in driving axion condensation, suggesting that the effective potential well could be non-trivially modified by the interplay between $\bar{m}$ and $\bar{f_a}$ within the Kerr-black-hole-cavity system \cite{Dolan:2012yt}.

\begin{figure*}[htbp]
	\begin{center}
		\includegraphics[width=0.43\linewidth]{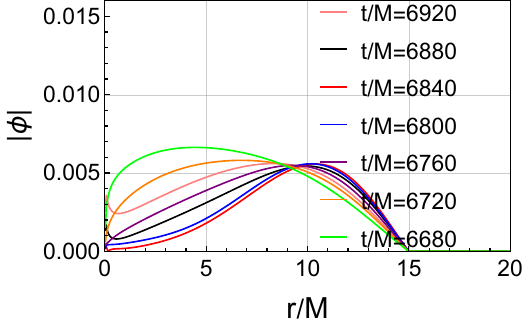}
        \includegraphics[scale=1.05]{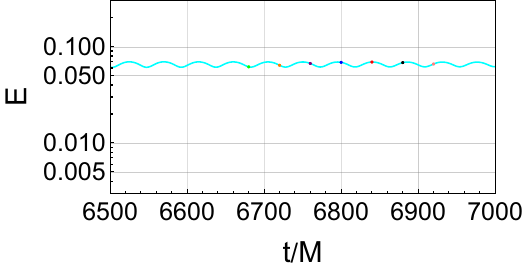}
	\end{center}
	\caption{Radial distribution of the axion field (left panel) for various time snapshots near the final state (right panel, color-coded) within a cavity of radius $r_{\text{mirr}}=15M$. The parameters are set to $\bar{q}=5, \bar{f_a}=0.1$, and $\bar{m}=0.4$.}
	\label{distribig4}
\end{figure*}

\begin{figure*}[htbp]
	\begin{center}
		\includegraphics[width=0.43\linewidth]{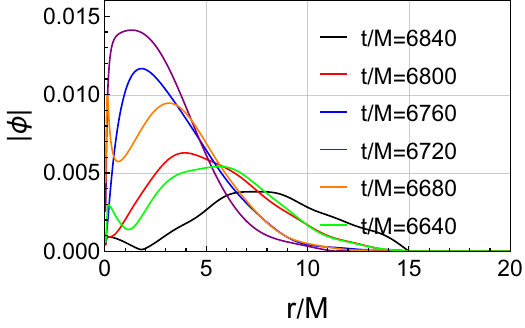}
        \includegraphics[scale=1.05]{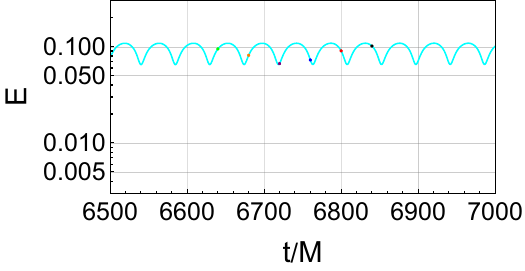}
	\end{center}
	\caption{Radial distribution of the axion field (left panel) for various time snapshots near the final state (right panel) within a cavity of radius $r_{\text{mirr}}=15M$. The parameters are set to $\bar{q}=5, \bar{f_a}=0.1$, and $\bar{m}=0.6$.}
	\label{distribig6}
\end{figure*}

Figures \ref{distribig4} and \ref{distribig6} depict the oscillatory dynamics for mass parameters $\bar{m}=0.4$ and $\bar{m}=0.6$, respectively. The temporal evolution, visualized through color-coding, reveals that the oscillation modes effectively localize the field, concentrating its density within $r \leq 15M$ and forming a coherent, dynamical axion cloud. When compared to the standard Kerr black hole scenario \cite{Yoshino:2012kn,Barranco:2011eyw}, which typically demands significantly higher computational resources, our cavity model confirms that while the natural potential well extends beyond the mirror radius $r_{\text{mirr}}$, the mirror-induced confinement provides a physically consistent approximation of the cloud's behavior. As the oscillations damp over time, the system with $\bar{m}=0.4$ is expected to relax into a quasi-stationary state. This confinement and condensation are even more pronounced for $\bar{m}=0.6$ (Fig. \ref{distribig6}), where the cloud oscillates in close proximity to the horizon. Peak density is achieved at specific instants corresponding to local minima of the total field energy at the extraction point. Although the selected snapshots do not span a single contiguous period, the strong correlation in energy values across different cycles allows us to reliably reconstruct the evolution of the cloud’s morphology.



\subsection{Bosenova}

\begin{figure}[htbp]
	\begin{center}
	   \includegraphics[width=0.8\linewidth]{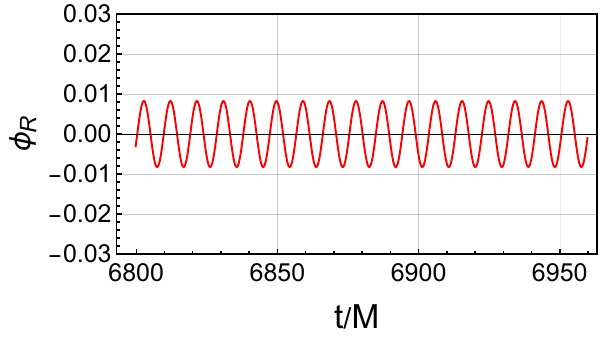}
       \includegraphics[width=0.8\linewidth]{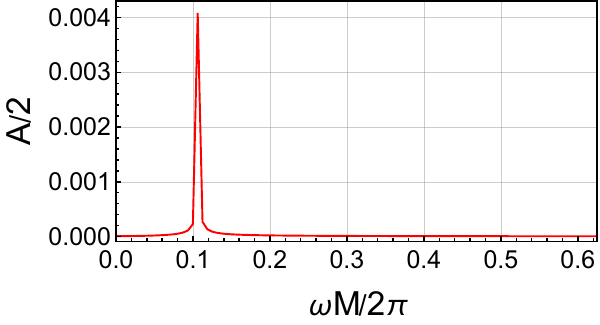}
	\end{center}
	\caption{Waveform of the axion and the corresponding frequency spectrum within a cavity of radius $r_{\text{mirr}}=15M$. The parameters are set to $\bar{f_a}=0.1, \bar{q}=5$, and $\bar{m}=0.2$.}
	\label{shorttimeM2}
\end{figure}

\begin{figure*}[htbp]
	\begin{center}
        \includegraphics[width=0.24\linewidth]{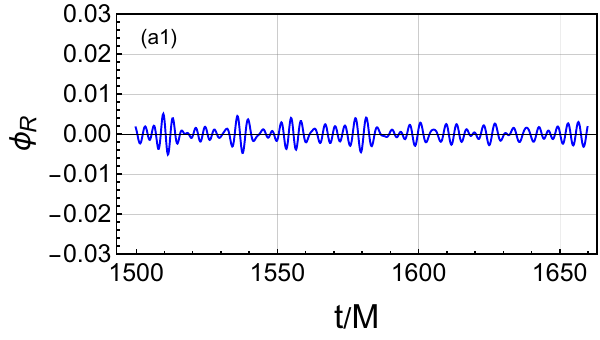}
        \includegraphics[width=0.24\linewidth]{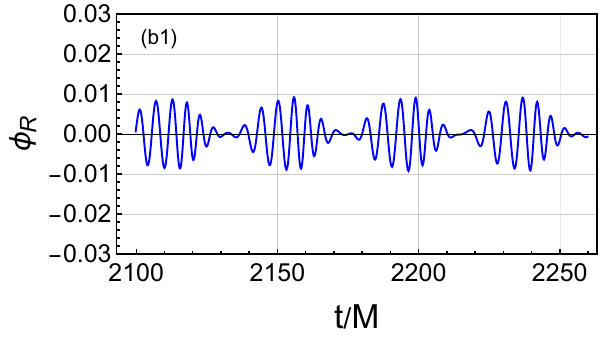}
        \includegraphics[width=0.24\linewidth]{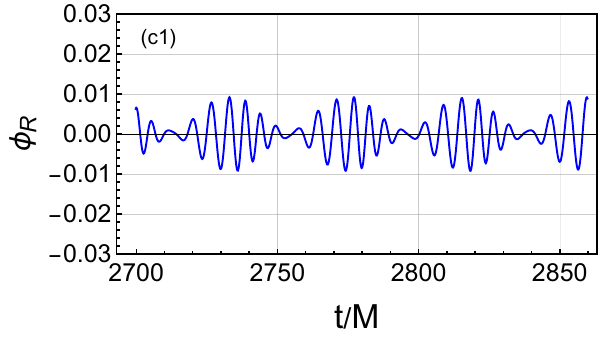}
        \includegraphics[width=0.24\linewidth]{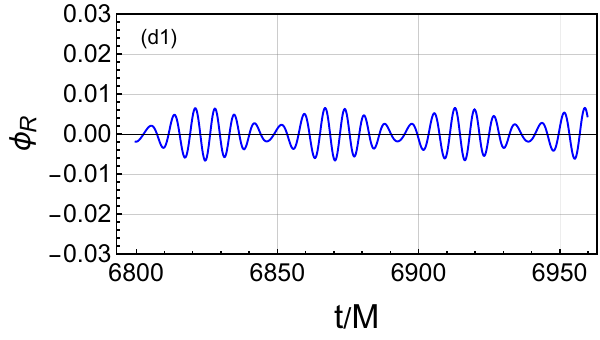}

        \includegraphics[width=0.24\linewidth]{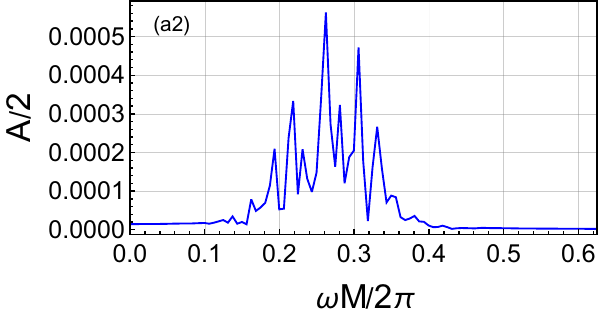}
        \includegraphics[width=0.24\linewidth]{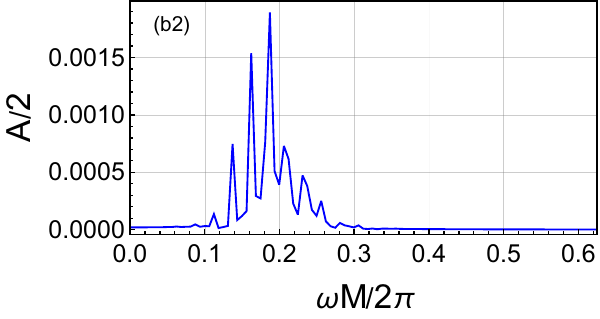}
        \includegraphics[width=0.24\linewidth]{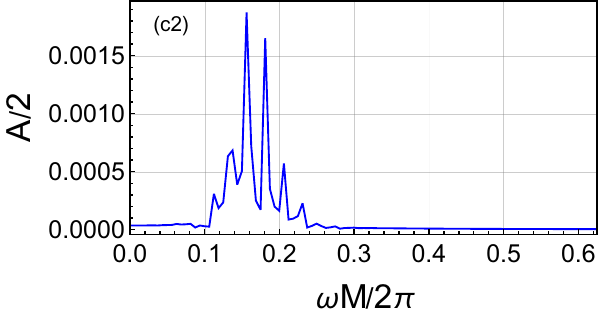}
        \includegraphics[width=0.24\linewidth]{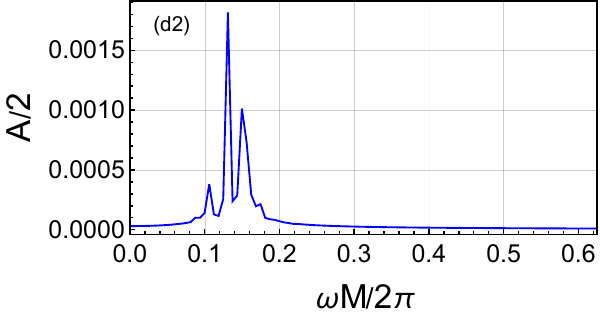}
	\end{center}
	\caption{Beating pattern in the axion field waveform and the corresponding frequency spectrum within a cavity of radius $r_{\text{mirr}}=15M$. The parameters are set to $\bar{f_a}=0.1, \bar{q}=5$, and $\bar{m}=0.4$.}
	\label{shorttimeM4}
\end{figure*}

\begin{figure*}[htbp]
	\begin{center}
        \includegraphics[width=0.24\linewidth]{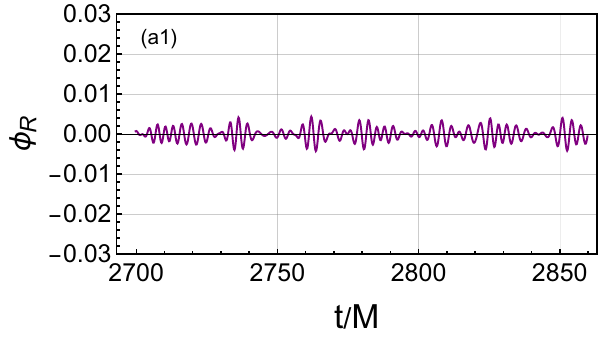}
        \includegraphics[width=0.24\linewidth]{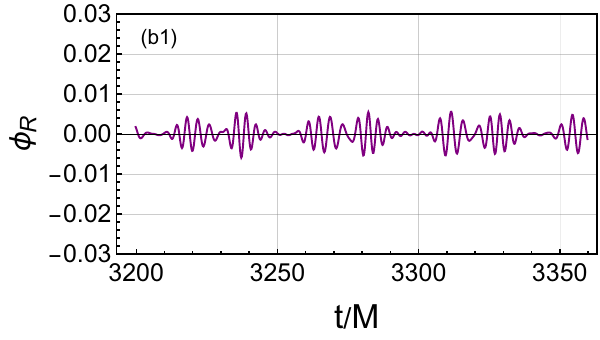}
        \includegraphics[width=0.24\linewidth]{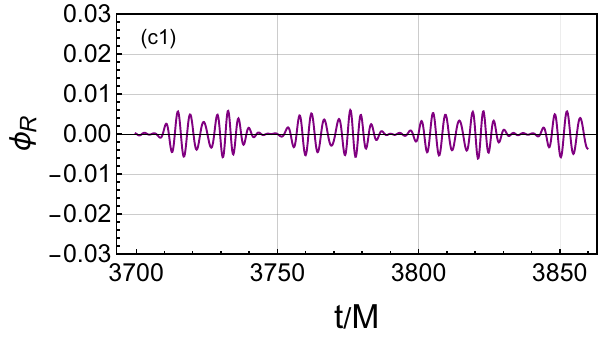}
        \includegraphics[width=0.24\linewidth]{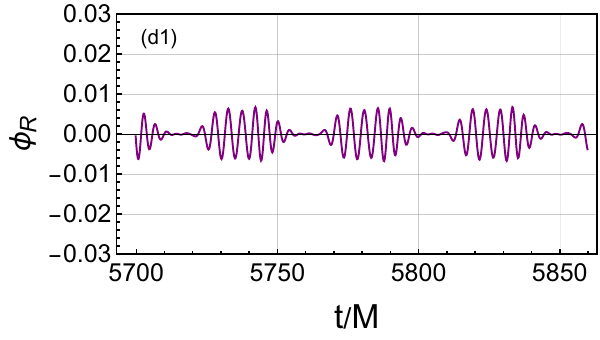}

        \includegraphics[width=0.24\linewidth]{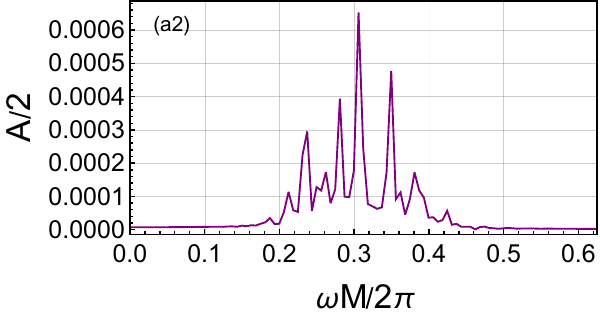}
        \includegraphics[width=0.24\linewidth]{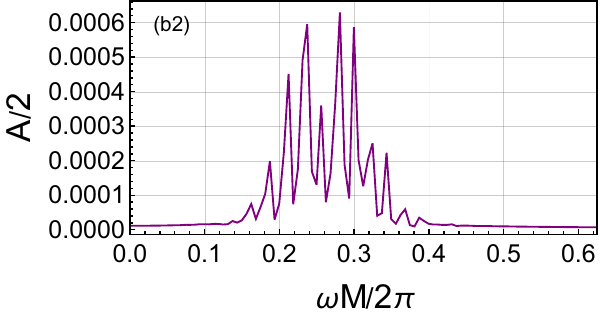}
        \includegraphics[width=0.24\linewidth]{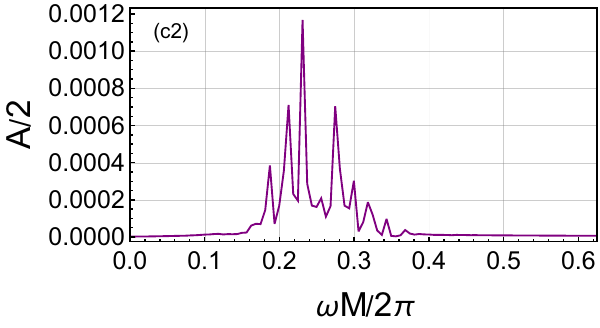}
        \includegraphics[width=0.24\linewidth]{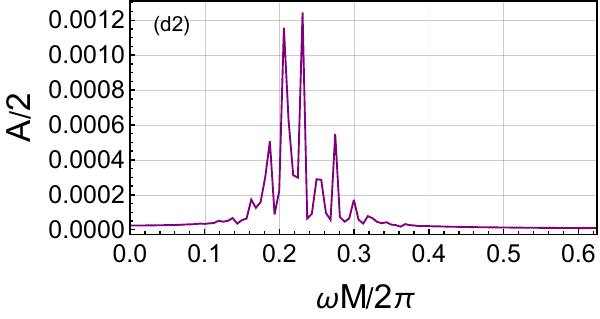}
	\end{center}
	\caption{Beating pattern in the axion field waveform and the corresponding frequency spectrum within a cavity of radius $r_{\text{mirr}}=15M$. The parameters are set to $\bar{f_a}=0.1, \bar{q}=5$, and $\bar{m}=0.6$.}
	\label{shorttimeM6}
\end{figure*}

To further elucidate the nonlinear evolution of the system, we further explore the Bosenova phenomenon of the axion reported in Ref. \cite{Yoshino:2012kn}. This process is usually triggered when the axion cloud, having grown exponentially through superradiance, reaches a critical energy where the attractive self-interactions modulated by the decay constant $\bar{f_a}$ cause the transition of the radial effective potential of the axion from the outer stable point to the inner stable point \cite{Yoshino:2012kn}, leading to a rapid non-linear collapse. By examining the field evolution within a refined temporal window, we identified distinct beating patterns \cite{Witek:2012tr, Okawa:2014nda}. These patterns and corresponding spectrums in Figs. \ref{shorttimeM2}, \ref{shorttimeM4}, and \ref{shorttimeM6} provide clear evidence for the coexistence of multiple excitation modes, a dynamical complexity that is typically absent in simple systems such as the RNBH.

Our numerical results confirm that Bosenova is triggered across several models, culminating in various saturation regimes dictated by the underlying parameters. The specific type and strength of these nonlinear instabilities can be characterized by the degree of energy overshooting \cite{Sanchis-Gual:2016tcm}. In the first panels of Figs. \ref{mirr10grp4} and \ref{mirr15grp4}, we observe that nonlinear modes, distinct from the fundamental mode, are rapidly re-absorbed by the black hole. Given the subsequent stability of the energy, we conclude that the system eventually becomes dominated by the fundamental mode. Notably, the case of $\bar{m}=0.6$ in subfigure (a) of Fig. \ref{mirr15grp4} reveals a specific configuration where the interplay between the mass parameter, decay constant, and mirror position effectively suppresses the Bosenova.

Furthermore, in the $\bar{f_a}=0.001$ case, we observe a higher level of residual energy following the Bosenova event when the mass parameter is increased. This suggests that a higher mass parameter facilitates the generation of additional quasi-stable nonlinear modes; although these modes do not dominate the total energy magnitude, they contribute to the long-term persistence of the field. The influence of the cavity boundary is also significant. Comparing the final two panels of Figs. \ref{mirr10grp4} and \ref{mirr15grp4} demonstrates that an extended mirror position ($r_{\text{mirr}}$) provides a larger spatial domain for the growth of nonlinear modes. This increased volume, combined with a prolonged superradiance timescale, allows for more substantial energy extraction. Finally, what need to be noticed is that Bosenova can happen in massless scenario once bosonic nonlinear backreaction to the spacetime is considered.

\begin{figure}
	\includegraphics[width=1\linewidth]{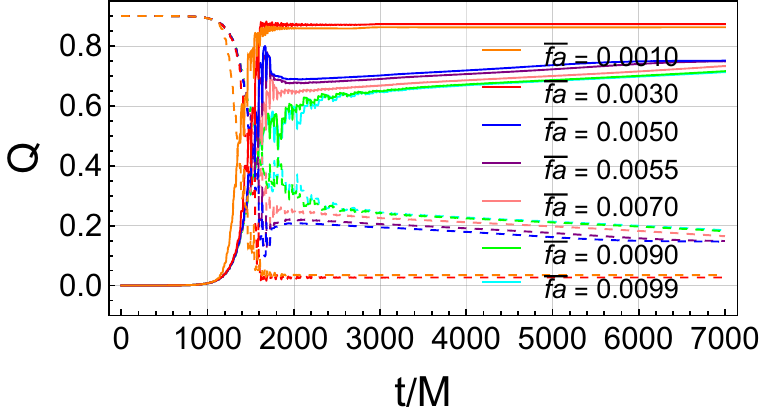}
	\caption{Charge transfer between the axion field and the black hole for various decay constants $\bar{f_a}$ within a cavity of radius $r_{\text{mirr}}=15M$, exhibiting a significant variation. Dashed and solid lines denote the charges of the black hole and the axion field, respectively. The parameters are set to $\bar{q}=5$ and $\bar{m}=0.4$.}
	\label{criticalfa}
\end{figure}

Before drawing final conclusions, we provide a brief discussion of the findings obtained thus far. Our results indicate that the influence of various parameters on the system's evolution is inherently non-linear and multi-faceted. Under diverse configurations, the conventional trends typically governed by a single parameter may diverge or even reverse. While exploring a vast parameter space may uncover individual correlations, a comprehensive understanding of the superradiant process ultimately demands a global perspective. Specifically, the interplay between the mass parameter $\bar{m}$ and the decay constant $\bar{f}_a$ arises from the formation of an effective potential well, shaped by both the scalar mass and the higher-order self-interaction terms. As suggested by Eq. \eqref{axionpotential}, different parameter combinations can lead to degenerate physical behaviors, where distinct sets of $\{\bar{m}, \bar{f_a}, r_{\text{mirr}}\}$ produce similar dynamical outcomes. This degeneracy implies that within certain overlapping regions of the parameter subspace, the system's evolution can be dominated by different physical mechanisms depending on the specific regime. However, the marginal sensitivity of these parameters is notable. As shown in Fig. \ref{criticalfa}, a distinct transition—or threshold—exists across different ranges of the decay constant. In this regime, the peak of the observed change signifies the charged bosonic field being pulled back toward the black hole. Beyond this threshold, all modes except the fundamental one are suppressed, marking a significant departure from the expected growth trends.

Beyond the standard parameter study, our simulations reveal rich phenomena that emerge under extreme physical conditions. Although some of these findings may deviate from commonly adopted astrophysical models, they remain of profound theoretical interest for understanding the limits of the Einstein-Maxwell-axion system. A particularly striking example is found in Figs. \ref{waveform} and \ref{waveformq}, where the irreducible mass of the black hole exhibits an anomalous reduction. While such a decrease is typically considered prohibited by the classical area theorem, its appearance here suggests the introduction of the mirror for the scalar field is unphysical.

\section{Conclusions} \label{sec:conc}

In this paper, we presented a comprehensive investigation into the superradiant evolution of a charged axion field around a RNBH, focusing on the intricate energy and charge exchange mechanisms within a cavity-enclosed system. By employing a controlled-variable approach across a broad parameter space, including the decay constant $\bar{f_a}$, field charge $\bar{q}$, scalar mass $\bar{m}$, and mirror radius $r_{\text{mirr}}$, we elucidated a relatively complete evolutionary picture of the ``hairy" black hole formation. Our findings reveal that the system's dynamics are not governed by isolated parameter dependencies but rather by a complex interplay of regimes. Notably, we identified a critical sensitivity threshold near $\bar{f_a} \sim 0.01$, beyond which the decay constant no longer exerts a dominant influence on the results.

The energy extraction process and the subsequent spatial morphology of the axion cloud are heavily modulated by the mass parameter and the proximity of the reflective boundary. We observed that while a closer boundary accelerates the superradiant extraction rate, it simultaneously suppresses the influence of other physical parameters and reduces the final energy yield. In the massive regime, the axion field exhibits rich condensation behavior and radial ``bounces," with larger mass parameters leading to stronger oscillations and clouds that reside closer to the black hole horizon. This behavior suggests that the potential well, shaped by the mass parameter and non-linear self-interaction, acts as the primary governor of the field's distribution, effectively mimicking the natural accumulation of axion cloud found in rotating black hole systems.

Beyond the standard evolutionary paths, our simulations highlighted a diverse range of dynamical endpoints, including stable saturation, slowly decaying oscillations, and long-term oscillatory regimes. These diverse outcomes emphasize that a global perspective is essential for understanding the superradiance process, as different parameter combinations within overlapping subspaces would trigger degenerate physical behaviors or lead to entirely new dynamical signatures. Looking forward, the high-density axion environments and the resulting hairy black hole configurations studied here provide a promising theoretical framework for the indirect detection of dark matter. The spatial distribution and temporal oscillations of the axion cloud hold significant implications for astrophysical observables, such as the dynamical evolution of black hole shadows, gravitational lensing effects, and the emission of superradiant gravitational waves. We anticipate that extending this research to more natural settings, such as Kerr backgrounds, and exploring a broader range of observation positions will further clarify the physics of these extreme regimes and serve as a valuable reference for future multi-messenger astronomical studies.

\acknowledgments

This work was supported in part by the National Natural Science Foundation of China (Grants No. 12575055 and No. 12247101), the Fundamental Research Funds for the Central Universities (Grant No. lzujbky-2025-jdzx07), the Natural Science Foundation of Gansu Province (No. 22JR5RA389, No.25JRRA799), the 111 Project under (Grant No. B20063). Yu-Peng Zhang was supported by ``Talent Scientific Fund of Lanzhou University".


\begin{thebibliography}{100}

\bibitem{Barack:2018yly}
L.~Barack, V.~Cardoso, S.~Nissanke, T.~P.~Sotiriou, A.~Askar, C.~Belczynski, G.~Bertone, E.~Bon, D.~Blas, and R.~Brito, \textit{et al.}
Class. Quant. Grav. \textbf{36}, 143001 (2019).

\bibitem{LIGOScientific:2016aoc}
B.~P.~Abbott \textit{et al.} [LIGO Scientific and Virgo],
Phys. Rev. Lett. \textbf{116}, 061102 (2016).

\bibitem{LIGOScientific:2016vlm}
B.~P.~Abbott \textit{et al.} [LIGO Scientific and Virgo],
Phys. Rev. Lett. \textbf{116}, 241102 (2016).

\bibitem{LIGOScientific:2016lio}
B.~P.~Abbott \textit{et al.} [LIGO Scientific and Virgo],
Phys. Rev. Lett. \textbf{116}, no.22, 221101 (2016)
[erratum: Phys. Rev. Lett. \textbf{121}, 129902 (2018)].

\bibitem{LIGOScientific:2016sjg}
B.~P.~Abbott \textit{et al.} [LIGO Scientific and Virgo],
Phys. Rev. Lett. \textbf{116}, 241103 (2016).

\bibitem{LIGOScientific:2019ysc}
B.~P.~Abbott \textit{et al.} [LIGO Scientific and Virgo],
Phys. Rev. D \textbf{100}, 064064 (2019).

\bibitem{LIGOScientific:2020stg}
R.~Abbott \textit{et al.} [LIGO Scientific and Virgo],
Phys. Rev. D \textbf{102}, 043015 (2020).

\bibitem{LIGOScientific:2017vwq}
B.~P.~Abbott \textit{et al.} [LIGO Scientific and Virgo],
Phys. Rev. Lett. \textbf{119}, 161101 (2017).

\bibitem{LIGOScientific:2018dkp}
B.~P.~Abbott \textit{et al.} [LIGO Scientific and Virgo],
Phys. Rev. Lett. \textbf{123}, 011102 (2019).

\bibitem{LIGOScientific:2018jsj}
B.~P.~Abbott \textit{et al.} [LIGO Scientific and Virgo],
Astrophys. J. Lett. \textbf{882}, L24 (2019).

\bibitem{LIGOScientific:2020kqk}
R.~Abbott \textit{et al.} [LIGO Scientific and Virgo],
Astrophys. J. Lett. \textbf{913}, L7 (2021).

\bibitem{KAGRA:2021duu}
R.~Abbott \textit{et al.} [KAGRA, VIRGO and LIGO Scientific],
Phys. Rev. X \textbf{13}, 011048 (2023).

\bibitem{EventHorizonTelescope:2019pgp}
K.~Akiyama \textit{et al.} [Event Horizon Telescope],
Astrophys. J. Lett. \textbf{875}, L1 (2019); \textbf{875},L2 (2019); \textbf{875}, L3 (2019); \textbf{875}, L4 (2019); \textbf{875}, L5 (2019); \textbf{875}, L6 (2019).

\bibitem{EventHorizonTelescope:2022wkp}
K.~Akiyama \textit{et al.} [Event Horizon Telescope],
Astrophys. J. Lett. \textbf{930}, L12 (2022); \textbf{930}, L13 (2022); \textbf{930}, L14 (2022); \textbf{930}, L15 (2022); \textbf{930}, L16 (2022); \textbf{930}, L17 (2022).

\bibitem{Falcke:1999pj}
H.~Falcke, F.~Melia, and E.~Agol,
Astrophys. J. Lett. \textbf{528}, L13 (2000).

\bibitem{Chen:2022scf}
S.~Chen, J.~Jing, W.-L.~Qian, and B.~Wang,
Sci. China Phys. Mech. Astron. \textbf{66}, 260401 (2023).

\bibitem{Penrose:1969pc}
R.~Penrose,
Riv. Nuovo Cim. \textbf{1}, 252 (1969).

\bibitem{Penrose:1971uk}
R.~Penrose and R.~M.~Floyd,
Nature \textbf{229}, 177 (1971).

\bibitem{Piran:1977dm}
T.~Piran and J.~Shaham,
Phys. Rev. D \textbf{16}, 1615 (1977).

\bibitem{Denardo:1973pyo}
G.~Denardo and R.~Ruffini,
Phys. Lett. B \textbf{45}, 259 (1973).

\bibitem{Wagh:1989zqa}
S.~M.~Wagh and N.~Dadhich,
Phys. Rept. \textbf{183}, 137 (1989).

\bibitem{Ruffini:2024dwq}
R.~Ruffini, M.~Prakapenia, H.~Quevedo, and S.~Zhang,
Phys. Rev. Lett. \textbf{134}, 081403 (2025).

\bibitem{Ruffini:2024irc}
R.~Ruffini, C.~L.~Bianco, M.~Prakapenia, H.~Quevedo, J.~A.~Rueda, and S.-R.~Zhang,
Phys. Rev. Res. \textbf{7}, 013203 (2025).

\bibitem{Blandford:1977ds}
R.~D.~Blandford and R.~L.~Znajek,
Mon. Not. Roy. Astron. Soc. \textbf{179}, 433 (1977).

\bibitem{Comisso:2020ykg}
L.~Comisso and F.~A.~Asenjo,
Phys. Rev. D \textbf{103}, 023014 (2021).

\bibitem{Wang:2022qmg}
C.-H.~Wang, C.-Q.~Pang, and S.-W.~Wei,
Phys. Rev. D \textbf{106}, 124050 (2022).

\bibitem{Zeldovich:1971mw}
Y.~B.~Zel'dovich,
Pis’ma Zh. Eksp. Teor. Fiz. \textbf{14} 270 (1971) [JETP Lett. \textbf{14}, 180 (1971)].


\bibitem{Zeldovich:1972zqp}
Y.~B.~Zel'dovich,
Zh. Eksp. Teor. Fiz \textbf{62} 2076 (1972)[Sov.Phys. JETP \textbf{35}, 1085 (1972)].


\bibitem{Teukolsky:1972my}
S.~A.~Teukolsky,
Phys. Rev. Lett. \textbf{29}, 1114 (1972).

\bibitem{Press:1972zz}
W.~H.~Press and S.~A.~Teukolsky,
Nature \textbf{238}, 211 (1972).

\bibitem{Brito:2015oca}
R.~Brito, V.~Cardoso, and P.~Pani,
Lect. Notes Phys. \textbf{906}, pp.1-237 (2015); Lect.Notes Phys. \textbf{971}, pp.1-293 (2020).

\bibitem{Bekenstein:1973mi}
J.~D.~Bekenstein,
Phys. Rev. D \textbf{7}, 949 (1973).

\bibitem{Zouros:1979iw}
T.~J.~M.~Zouros and D.~M.~Eardley,
Annals Phys. \textbf{118}, 139 (1979).


\bibitem{Detweiler:1980uk}
S.~L.~Detweiler,
Phys. Rev. D \textbf{22}, 2323 (1980).

\bibitem{Cardoso:2004nk}
V.~Cardoso, O.~J.~C.~Dias, J.~P.~S.~Lemos, and S.~Yoshida,
Phys. Rev. D \textbf{70}, 044039 (2004)
[erratum: Phys. Rev. D \textbf{70}, 049903 (2004)].

\bibitem{Dolan:2007mj}
S.~R.~Dolan,
Phys. Rev. D \textbf{76}, 084001 (2007).

\bibitem{RosaJ}
Rosa.~J,

JHEP \textbf{06}, 015 (2010).


\bibitem{Dolan:2012yt}
S.~R.~Dolan,
Phys. Rev. D \textbf{87}, 124026 (2013).

\bibitem{Witek:2012tr}
H.~Witek, V.~Cardoso, A.~Ishibashi, and U.~Sperhake,
Phys. Rev. D \textbf{87}, 043513 (2013).

\bibitem{Pani:2012vp}
P.~Pani, V.~Cardoso, L.~Gualtieri, E.~Berti, and A.~Ishibashi,
Phys. Rev. Lett. \textbf{109}, 131102 (2012).

\bibitem{Sanchis-Gual:2015lje}
N.~Sanchis-Gual, J.~C.~Degollado, P.~J.~Montero, J.~A.~Font, and C.~Herdeiro,
Phys. Rev. Lett. \textbf{116}, 141101 (2016).

\bibitem{Sanchis-Gual:2016tcm}
N.~Sanchis-Gual, J.~C.~Degollado, C.~Herdeiro, J.~A.~Font, and P.~J.~Montero,
Phys. Rev. D \textbf{94}, no.4, 044061 (2016).

\bibitem{Herdeiro:2013pia}
C.~A.~R.~Herdeiro, J.~C.~Degollado, and H.~F.~R{\'u}narsson,
Phys. Rev. D \textbf{88}, 063003 (2013).

\bibitem{Hod:2012wmy}
S.~Hod,
Phys. Lett. B \textbf{713}, 505 (2012).


\bibitem{Dolan:2015dha}
S.~R.~Dolan, S.~Ponglertsakul, and E.~Winstanley,
Phys. Rev. D \textbf{92}, 124047 (2015).


\bibitem{Sanchis-Gual:2014ewa}
N.~Sanchis-Gual, J.~C.~Degollado, P.~J.~Montero, and J.~A.~Font,
Phys. Rev. D \textbf{91}, 043005 (2015).

\bibitem{Herdeiro:2014goa}
C.~A.~R.~Herdeiro and E.~Radu,
Phys. Rev. Lett. \textbf{112}, 221101 (2014).

\bibitem{Zilhao:2015tya}
M.~Zilh{\~a}o, H.~Witek, and V.~Cardoso,
Class. Quant. Grav. \textbf{32}, 234003 (2015).

\bibitem{Herdeiro:2016tmi}
C.~Herdeiro, E.~Radu, and H.~R{\'u}narsson,
Class. Quant. Grav. \textbf{33}, 154001 (2016).

\bibitem{Ganchev:2017uuo}
B.~Ganchev and J.~E.~Santos,
Phys. Rev. Lett. \textbf{120}, 171101 (2018).

\bibitem{Herdeiro:2017phl}
C.~A.~R.~Herdeiro and E.~Radu,
Phys. Rev. Lett. \textbf{119}, 261101 (2017).

\bibitem{Witek:2010qc}
H.~Witek, V.~Cardoso, C.~Herdeiro, A.~Nerozzi, U.~Sperhake, and M.~Zilhao,
Phys. Rev. D \textbf{82}, 104037 (2010).

\bibitem{Cardoso:2012qm}
V.~Cardoso, L.~Gualtieri, C.~Herdeiro, U.~Sperhake, P.~M.~Chesler, L.~Lehner, S.~C.~Park, H.~S.~Reall, C.~F.~Sopuerta, and D.~Alic, \textit{et al.}
Class. Quant. Grav. \textbf{29}, 244001 (2012).

\bibitem{Okawa:2014nda}
H.~Okawa, H.~Witek, and V.~Cardoso,
Phys. Rev. D \textbf{89}, 104032 (2014).

\bibitem{Okawa:2015fsa}
H.~Okawa,
Class. Quant. Grav. \textbf{32}, 214003 (2015).

\bibitem{Bosch:2016vcp}
P.~Bosch, S.~R.~Green, and L.~Lehner,
Phys. Rev. Lett. \textbf{116}, 141102 (2016).

\bibitem{East:2017ovw}
W.~E.~East and F.~Pretorius,
Phys. Rev. Lett. \textbf{119}, 041101 (2017).

\bibitem{East:2018glu}
W.~E.~East,
Phys. Rev. Lett. \textbf{121}, 131104 (2018).

\bibitem{Baryakhtar:2020gao}
M.~Baryakhtar, M.~Galanis, R.~Lasenby, and O.~Simon,
Phys. Rev. D \textbf{103}, 095019 (2021).

\bibitem{Zhang:2023qtn}
C.-Y.~Zhang, Q.~Chen, Y.~Liu, Y.~Tian, B.~Wang, and H.~Zhang,
Phys. Rev. D \textbf{110}, L041505 (2024).

\bibitem{Zhang:2025jlb}
C.-Y.~Zhang, Z.~Zhang, and R.~Zheng,
Sci. China Phys. Mech. Astron. \textbf{68}, 250411 (2025).

\bibitem{Saffin:2022tub}
P.~M.~Saffin, Q.-X.~Xie, and S.-Y.~Zhou,
Phys. Rev. Lett. \textbf{131}, 11 (2023).

\bibitem{Gao:2023gof}
H.-Y.~Gao, P.~M.~Saffin, Y.-J.~Wang, Q.-X.~Xie, and S.-Y.~Zhou,
Sci. China Phys. Mech. Astron. \textbf{67}, 260413 (2024).

\bibitem{Zhou:2023sps}
L.~Zhou, R.~Brito, Z.-F.~Mai, and L.~Shao,
Phys. Rev. D \textbf{108}, 103025 (2023).

\bibitem{Luo:2024gqo}
Z.-H.~Luo and Y.-L.~Zhang,
Eur. Phys. J. C \textbf{85}, 507 (2025).

\bibitem{Jha:2022tdl}
S.~K.~Jha, M.~Khodadi, A.~Rahaman, and A.~Sheykhi,
Phys. Rev. D \textbf{107}, 084052 (2023).

\bibitem{Piovella:2023aou}
N.~Piovella and S.~Olivares,
Symmetry \textbf{15}, 1817 (2023).

\bibitem{Dai:2023zcj}
D.-C.~Dai and D.~Stojkovic,
Phys. Rev. D \textbf{108}, 084024 (2023).

\bibitem{Dai:2023ewf}
D.-C.~Dai and D.~Stojkovic,
Phys. Lett. B \textbf{843}, 138056 (2023).

\bibitem{Karmakar:2023hlb}
R.~Karmakar and D.~Maity,
Eur. Phys. J. C \textbf{85}, 1191 (2025).

\bibitem{Peccei:1977hh}
R.~D.~Peccei and H.~R.~Quinn,
Phys. Rev. Lett. \textbf{38}, 1440 (1977).

\bibitem{Marsh:2015xka}
D.~J.~E.~Marsh,
Phys. Rept. \textbf{643}, 1 (2016).

\bibitem{Kodama:2011zc}
H.~Kodama and H.~Yoshino,
Int. J. Mod. Phys. Conf. Ser. \textbf{7}, 84 (2012).

\bibitem{Yoshino:2013ofa}
H.~Yoshino and H.~Kodama,
PTEP \textbf{2014}, 043E02 (2014).

\bibitem{Yoshino:2015nsa}
H.~Yoshino and H.~Kodama,
Class. Quant. Grav. \textbf{32}, 214001 (2015).

\bibitem{Yoshino:2012kn}
H.~Yoshino and H.~Kodama,
Prog. Theor. Phys. \textbf{128}, 153 (2012).

\bibitem{Blas:2020nbs}
D.~Blas and S.~J.~Witte,
Phys. Rev. D \textbf{102}, 103018 (2020).

\bibitem{Branco:2023frw}
N.~P.~Branco, R.~Z.~Ferreira, and J.~G.~Rosa,
JCAP \textbf{04}, 003 (2023).

\bibitem{Omiya:2022gwu}
H.~Omiya, T.~Takahashi, T.~Tanaka, and H.~Yoshino,
JCAP \textbf{06}, 016 (2023).

\bibitem{Takahashi:2021yhy}
T.~Takahashi, H.~Omiya, and T.~Tanaka,
PTEP \textbf{2022}, 043E01 (2022).

\bibitem{Takahashi:2023flk}
T.~Takahashi, H.~Omiya, and T.~Tanaka,
Phys. Rev. D \textbf{107}, 103020 (2023).

\bibitem{Takahashi:2021eso}
T.~Takahashi and T.~Tanaka,
JCAP \textbf{10}, 031 (2021).

\bibitem{Yang:2017lpm}
Q.~Yang, L.-W.~Ji, B.~Hu, Z.-J.~Cao, and R.-G.~Cai,
Res. Astron. Astrophys. \textbf{18}, 065 (2018).

\bibitem{Fukuda:2019ewf}
H.~Fukuda and K.~Nakayama,
JHEP \textbf{01}, 128 (2020).

\bibitem{Filippini:2019cqk}
F.~Filippini and G.~Tasinato,
Class. Quant. Grav. \textbf{36}, 215015 (2019).

\bibitem{Banerjee:2019xds}
I.~Banerjee, S.~Sau, and S.~SenGupta,
Phys. Rev. D \textbf{101}, 104057 (2020).

\bibitem{Choudhary:2020pxy}
S.~Choudhary, N.~Sanchis-Gual, A.~Gupta, J.~C.~Degollado, S.~Bose, and J.~A.~Font,
Phys. Rev. D \textbf{103}, 044032 (2021).

\bibitem{Delgado:2020hwr}
J.~F.~M.~Delgado, C.~A.~R.~Herdeiro, and E.~Radu,
Phys. Rev. D \textbf{103}, 104029 (2021).

\bibitem{Zhang:2022rex}
Y.-P.~Zhang, M.~Gracia-Linares, P.~Laguna, D.~Shoemaker, and Y.-X.~Liu,
Phys. Rev. D \textbf{107}, 044039 (2023).

\bibitem{Bamber:2022pbs}
J.~Bamber, J.~C.~Aurrekoetxea, K.~Clough, and P.~G.~Ferreira,
Phys. Rev. D \textbf{107}, 024035 (2023).

\bibitem{Herdeiro:2023roz}
C.~A.~R.~Herdeiro and E.~Radu,
Phys. Rev. Lett. \textbf{131}, 121401 (2023).

\bibitem{Leong:2023nuk}
S.~H.~W.~Leong, J.~Calder{\'o}n Bustillo, M.~Gracia-Linares, and P.~Laguna,
Phys. Rev. D \textbf{108}, 124079 (2023).

\bibitem{Aurrekoetxea:2023jwk}
J.~C.~Aurrekoetxea, K.~Clough, J.~Bamber, and P.~G.~Ferreira,
Phys. Rev. Lett. \textbf{132}, 211401 (2024).

\bibitem{Aurrekoetxea:2024cqd}
J.~C.~Aurrekoetxea, J.~Marsden, K.~Clough, and P.~G.~Ferreira,
Phys. Rev. D \textbf{110}, 083011 (2024).

\bibitem{Guerra:2019srj}
D.~Guerra, C.~F.~B.~Macedo, and P.~Pani,
JCAP \textbf{09}, 061 (2019)
[erratum: JCAP \textbf{06}, E01 (2020)].

\bibitem{Delgado:2020udb}
J.~F.~M.~Delgado, C.~A.~R.~Herdeiro, and E.~Radu,
JCAP \textbf{06}, 037 (2020).

\bibitem{Zeng:2021oez}
Y.-B.~Zeng, S.-Y.~Cui, H.-B.~Li, S.-X.~Sun, Y.-P.~Zhang, and Y.-Q.~Wang,
Eur. Phys. J. C \textbf{84}, 187 (2024).

\bibitem{Zeng:2023hvq}
Y.-B.~Zeng, S.-X.~Sun, S.-Y.~Cui, Y.-P.~Zhang, and Y.-Q.~Wang,
[arXiv:2309.05743 [gr-qc]].

\bibitem{Sakurai:2023hkg}
Y.~Sakurai, C.~M.~Yoo, A.~Naruko, and D.~Yamauchi,
JCAP \textbf{04}, 033 (2024).

\bibitem{Spieksma:2023vwl}
T.~F.~M.~Spieksma, E.~Cannizzaro, T.~Ikeda, V.~Cardoso, and Y.~Chen,
Phys. Rev. D \textbf{108}, 063013 (2023).

\bibitem{Boskovic:2018lkj}
M.~Boskovic, R.~Brito, V.~Cardoso, T.~Ikeda, and H.~Witek,
Phys. Rev. D \textbf{99}, 035006 (2019).

\bibitem{Caputo:2024oqc}
A.~Caputo and G.~Raffelt,
PoS \textbf{COSMICWISPers}, 041 (2024).

\bibitem{Caputo:2025oap}
A.~Caputo, G.~Franciolini, and S.~J.~Witte,
[arXiv:2507.21788 [hep-ph]].

\bibitem{Arvanitaki:2014wva}
A.~Arvanitaki, M.~Baryakhtar, and X.~Huang,
Phys. Rev. D \textbf{91}, 084011 (2015).

\bibitem{Arvanitaki:2016qwi}
A.~Arvanitaki, M.~Baryakhtar, S.~Dimopoulos, S.~Dubovsky, and R.~Lasenby,
Phys. Rev. D \textbf{95}, 043001 (2017).


\bibitem{Dorlis:2025zzz}
P.~Dorlis, N.~E.~Mavromatos, S.~Sarkar, and S.~N.~Vlachos,
Phys. Rev. Lett. \textbf{135}, 151501 (2025).

\bibitem{Dorlis:2025amf}
P.~Dorlis, N.~E.~Mavromatos, S.~Sarkar, and S.~N.~Vlachos,
[arXiv:2507.23475 [gr-qc]].

\bibitem{Mavromatos:2025ofn}
N.~E.~Mavromatos, P.~Dorlis, S.~Sarkar, and S.~N.~Vlachos,
[arXiv:2512.14951 [gr-qc]].

\bibitem{Cordero-Carrion:2012qac}
I.~Cordero-Carrion and P.~Cerda-Duran,
[arXiv:1211.5930 [math-ph]].

\bibitem{Baumgarte:2012xy}
T.~W.~Baumgarte, P.~J.~Montero, I.~Cordero-Carrion, and E.~Muller,
Phys. Rev. D \textbf{87}, 044026 (2013).

\bibitem{Montero:2012yr}
P.~J.~Montero and I.~Cordero-Carrion,
Phys. Rev. D \textbf{85}, 124037 (2012).

\bibitem{Alcubierre:2011pkc}
M.~Alcubierre and M.~D.~Mendez,
Gen. Rel. Grav. \textbf{43}, 2769 (2011).

\bibitem{GrillidiCortona:2015jxo}
G.~Grilli di Cortona, E.~Hardy, J.~Pardo Vega, and G.~Villadoro,
JHEP \textbf{01}, 034 (2016).

\bibitem{Bona:1997hp}
C.~Bona, J.~Masso, E.~Seidel, and J.~Stela,
Phys. Rev. D \textbf{56}, 3405 (1997).

\bibitem{Alcubierre:2002kk}
M.~Alcubierre, B.~Bruegmann, P.~Diener, M.~Koppitz, D.~Pollney, E.~Seidel, and R.~Takahashi,
Phys. Rev. D \textbf{67}, 084023 (2003).

\bibitem{Zhang:2023qxf}
Y.-P.~Zhang, S.-X.~Sun, Y.-Q.~Wang, S.-W.~Wei, P.~Laguna, and Y.-X.~Liu,
Phys. Rev. Res. \textbf{6}, 033187 (2024).

\bibitem{Zhang:2024wci}
Y.-P.~Zhang, S.-J.~Yang, S.-W.~Wei, W.-D.~Guo, and Y.-X.~Liu,
Phys. Rev. D \textbf{111}, 104005 (2025).

\bibitem{Alcubierre:2009ij}
M.~Alcubierre, J.~C.~Degollado, and M.~Salgado,
Phys. Rev. D \textbf{80}, 104022 (2009).

\bibitem{Corelli:2021ikv}
F.~Corelli, T.~Ikeda, and P.~Pani,
Phys. Rev. D \textbf{104}, 084069 (2021).

\bibitem{Torres:2014fga}
J.~M.~Torres and M.~Alcubierre,
Gen. Rel. Grav. \textbf{46}, 1773 (2014).

\bibitem{Barranco:2011eyw}
J.~Barranco, A.~Bernal, J.~C.~Degollado, A.~Diez-Tejedor, M.~Megevand, M.~Alcubierre, D.~Nunez, and O.~Sarbach,
Phys. Rev. D \textbf{84}, 083008 (2011).

\end{thebibliography}
\end{document}